\documentclass{PoS}

\def\lsim{\,\lower2truept\hbox{${< \atop\hbox{\raise4truept\hbox{$\sim$}}}$}\,}
\def\gsim{\,\lower2truept\hbox{${> \atop\hbox{\raise4truept\hbox{$\sim$}}}$}\,}

\title{Polarized synchrotron emission}

\ShortTitle{Polarized synchrotron emission}

\author{\speaker{Carlo Burigana}
\\
        INAF-IASF Bologna\\
        Via Gobetti 101, I-40129, Bologna, Italy\\
        E-mail: \email{burigana@iasfbo.inaf.it}}

\author{Laura La Porta\thanks{Member of the
          International Max Planck Research Shool (IMPRS)
for Radio and Infrared Astronomy at the Universities
           of Bonn and Cologne.}, Wolfgang Reich, Patricia Reich\\
        Max-Planck-Institut f\"ur Radioastronomie\\
        Auf dem H\"ugel, 69, D-53121 Bonn, Germany\\
        E-mail: \email{laporta@mpifr-bonn.mpg.de, wreich@mpifr-bonn.mpg.de, preich@mpifr-bonn.mpg.de}}

\author{Joaquin Gonzalez-Nuevo, Marcella Massardi\\
        SISSA/ISAS\\
        Via Beirut 2-4, I-34014 Trieste, Italy\\
        E-mail: \email{gnuevo@sissa.it, massardi@sissa.it}}

\author{Gianfranco De Zotti\\
        INAF - Osservatorio Astronomico di Padova\\
        Vicolo dell'Osservatorio 5, I-35122 Padova, Italy\\
        E-mail: \email{dezotti@pd.astro.it}}

\abstract{Galactic synchrotron emission represents the most relevant 
foreground contamination in cosmic microwave background (CMB) anisotropy observations 
at angular scales $\theta \gsim 1^\circ$ and frequencies $\nu \lsim 70$~GHz.
The accurate understanding 
of its polarization properties
is crucial to extract the cosmological information
contained in the CMB polarization anisotropy.
Radio surveys at $\nu \sim 1$~GHz offer the unique opportunity to study 
Galactic synchrotron emission where it represents the dominant 
component, possibly except for regions close to the Galactic plane
where free-free emission is also important. We review the 
observational status of Galactic radio surveys at scales $\theta \gsim 0.5^\circ$.
Leiden surveys, thanks to their frequency coverage from 0.408~GHz
to 1.411~GHz, still remain of fundamental importance 
for the comprehension of depolarization phenomena.
Recent surveys at 1.42~GHz (in both total intensity and polarization)
with a better sensitivity and sky sampling now cover both celestial hemispheres and 
allow to accurately map the correlation properties of the diffuse synchrotron emission.
We present an analysis of these surveys in terms of angular power spectrum.
A comparison of a simple frequency extrapolation of these results
with the recent WMAP results shows that we are close to map the bulk
of the diffuse synchrotron polarization fluctuations and to understand 
the corresponding implications for CMB experiments.}

\FullConference{CMB and Physics of the Early Universe\\
		 20-22 April 2006\\
		 Ischia, Italy}

\begin{document}

\section{Introduction}

The Galactic polarized diffuse synchrotron radiation is expected
to play the major role at frequencies below 70~GHz on intermediate
and large angular scales ($\theta \gsim 30'$), 
at least at medium and high Galactic latitudes where satellites have the 
clearest view of the CMB anisotropies.
At about 1~GHz the synchrotron emission is the most important radiative
mechanism out of the Galactic plane, while at low latitudes it is 
comparable with the bremsstrahlung; 
however the free-free emission is 
unpolarized, whereas the 
synchrotron radiation could reach a theoretical intrinsic
degree of polarization of about $75\%$ (see \cite{ginzburg65}). 
Consequently, radio frequencies are the natural range for studying it
(see \cite{reich06}), 
though it might be affected by Faraday rotation and depolarization.

It is standard practice to characterize the CMB anisotropies  
in terms of angular power spectrum
(\cite{peebles,kamion,zald})
(APS)
as a function of the multipole, $\ell$ (inversely proportional to 
the angular scale, $\ell \simeq {180}/{\theta(^{\circ})}$),
in total intensity ($T$ mode), polarization ($PI$, 
$E$, and $B$ modes), and cross-correlation modes
(in particular, we will consider the $TE$ mode).
Although the APS is certainly not exhaustive to fully characterize 
the complexity of the 
intrinsically non-Gaussian Galactic emission,
it is of increasing usage also to characterize 
the Galactic foreground correlation properties
in both data analysis and theoretical studies
(see e.g. \cite{bacci01,chepurnov,cho_lazarian}).
Using the {\tt anafast} facility of 
the {\tt HEALPix}~\footnote{http://healpix.jpl.nasa.gov/}
package by \cite{gorski05}, we computed the 
APS of the above modes for the various considered surveys, sky areas,
and frequencies.

\section{Analysis of the Leiden polarization surveys}

\begin{figure}[t]
   \centering
   \begin{tabular}{ccc}
\includegraphics[width=2.8cm,angle=90,clip=]{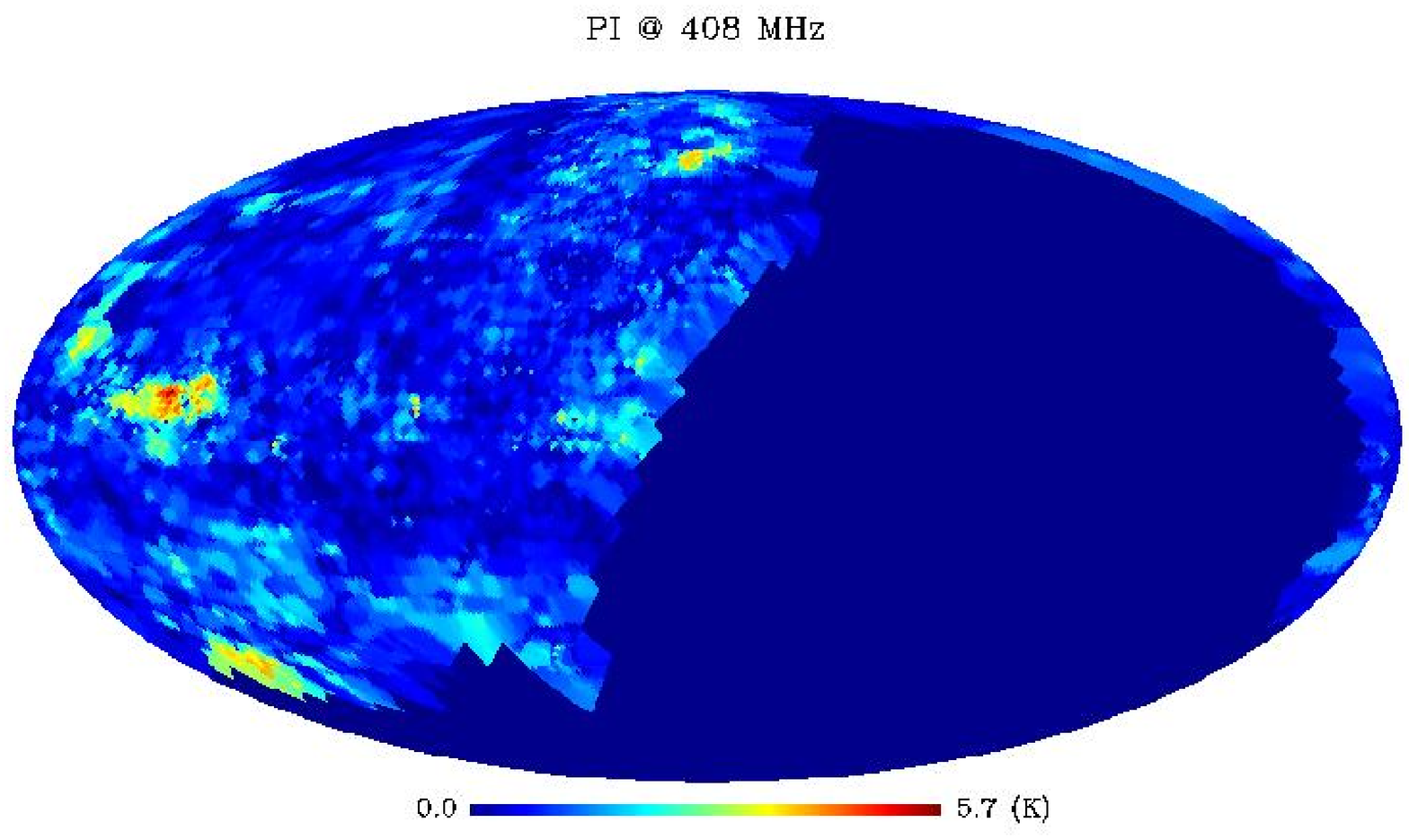}&
\includegraphics[width=2.8cm,angle=90,clip=]{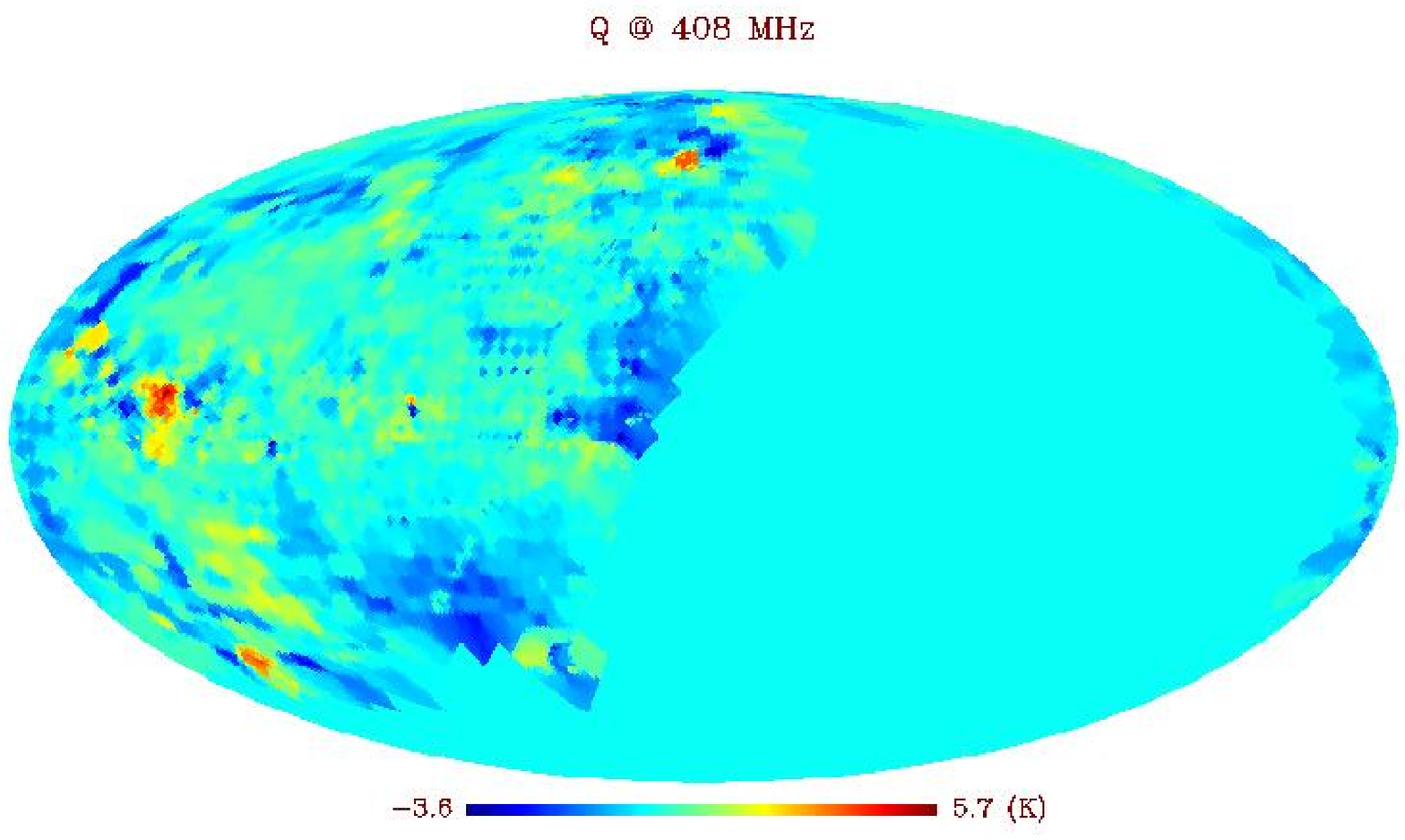}&
\includegraphics[width=2.8cm,angle=90,clip=]{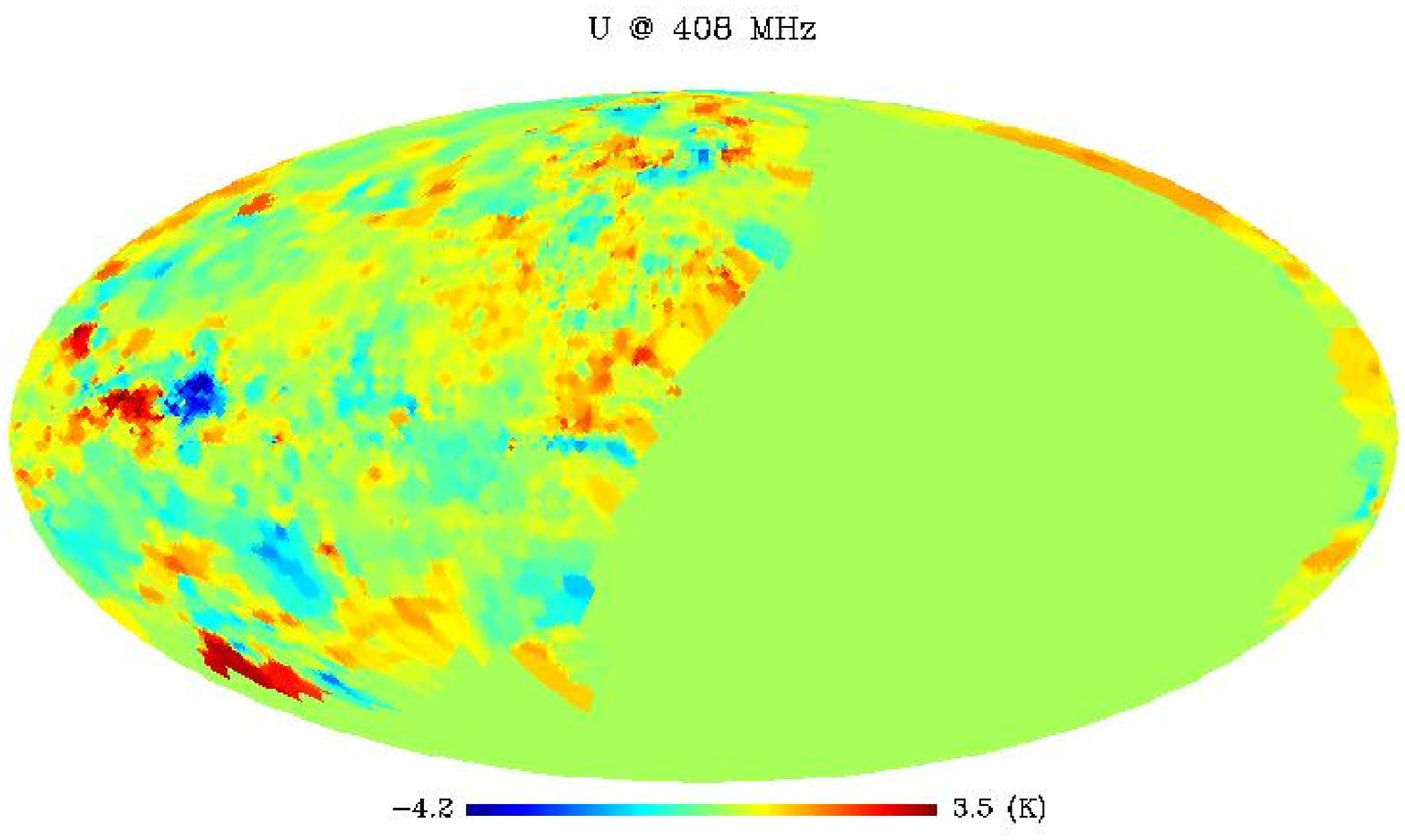}\\
\includegraphics[width=2.8cm,angle=90,clip=]{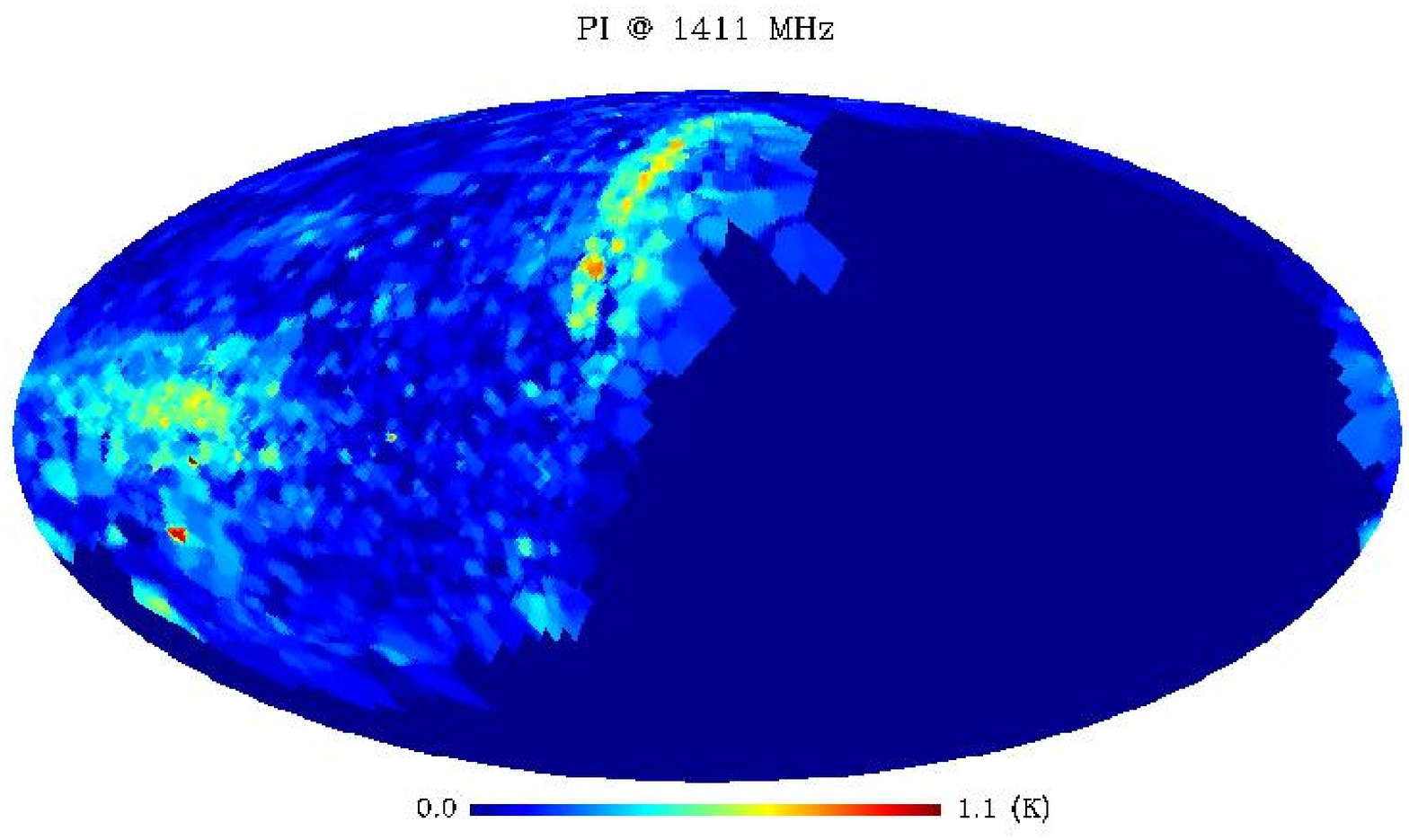}&
\includegraphics[width=2.8cm,angle=90,clip=]{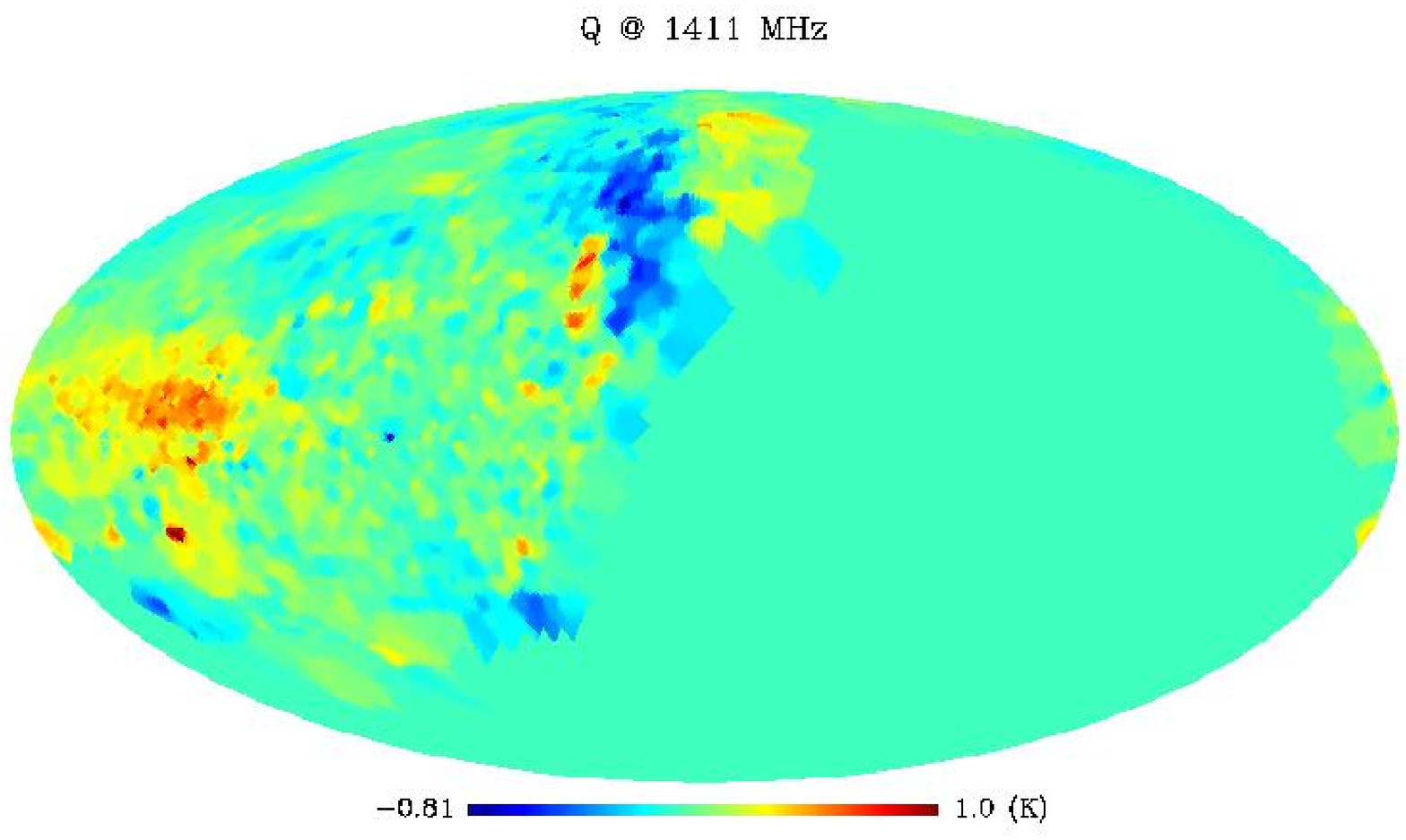}&
\includegraphics[width=2.8cm,angle=90,clip=]{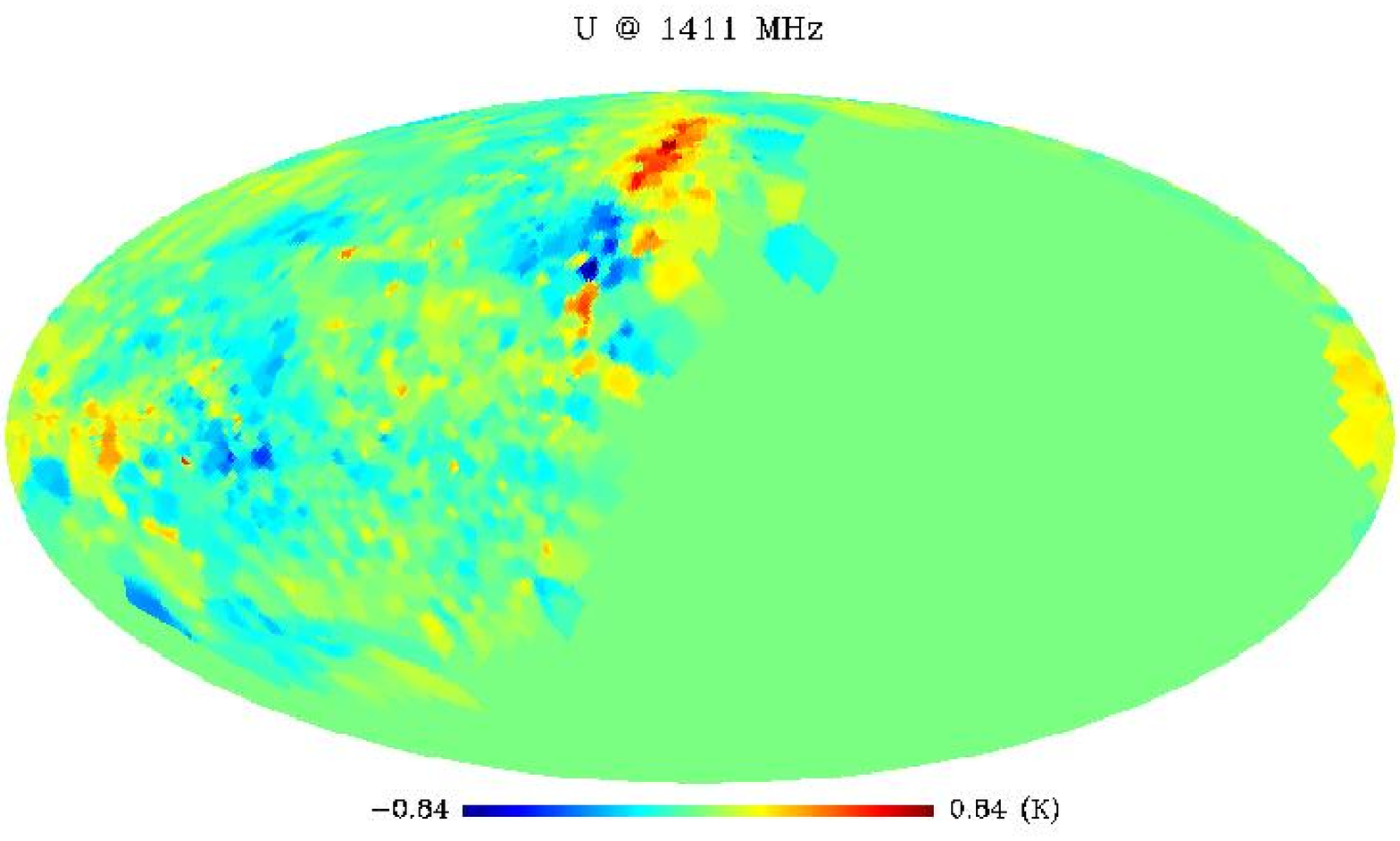}\\
   \end{tabular}
   \caption{$PI$, $Q$, and $U$ maps at 408~MHz and 1411~MHz 
obtained from the Leiden surveys data tables. Adapted from \cite{laportaburigana06}.}
   \label{allmaps}
 \end{figure}

Nowadays the only available radio data 
suitable for multifrequency studies 
on large scales are the so called Leiden surveys
(\cite{spo76}),
globally covering a sky fraction of about $\sim 50 \%$,
mainly in the northern Galactic hemisphere.
These linear polarization surveys are the result 
of different observational campaigns carried out 
in the sixties 
with the Dwingeloo 25-m radio telescope
at 408, 465, 610, 820 and 1411~MHz with angular resolutions
respectively of $\theta_{HPBW} = 2.3^{\circ}, 2.0^{\circ}, 
1.5^{\circ}, 1.0^{\circ}, 0.6^{\circ}$.
The complete data sets
as well as the observations reduction and calibration
methods have been presented by \cite{spo76}.
Specific interpolation methods
to project these surveys into {\tt HEALPix} maps
with pixel size $\simeq 0.92^{\circ}$ 
(see Fig.~\ref{allmaps})
have been recently implemented (\cite{laportaburigana06}),
catching with the main problem of these surveys, i.e.
their poor average sampling across the sky and their limited sensitivity
($\sim 0.34, 0.33, 0.16, 0.11$ and $0.06$~K from
the lower to the higher frequency, respectively).
Such methods have been successfully tested against simulations.

Three sky areas
with sampling significantly better than the average
(by a factor $\simeq 4$; \cite{bacci01,buriganalaporta02,laportaburigana06})
permit to reach multipoles $\ell \simeq 100$:
patch 1 [($110^\circ \le l \le 160^\circ$, $0^\circ \le b \le 20^\circ$)];
patch 2 [($5^\circ \le l \le 80^\circ$, $b \ge 50^\circ$) together with
($0^\circ \le l \le 5^\circ$, $b \ge 60^\circ$) and
($335^\circ \le l \le 360^\circ$, $b \ge 60^\circ$)];
patch 3 [($10^\circ \le l \le 80^\circ$, $b \ge 70^\circ$)]. 
These patches 
are associated with the brightest structures
of the polarized radio sky, i.e. the Fan Region (patch 1) and the
North Polar Spur (NPS) (patch 2 and 3). The NPS
has been extensively studied (see \cite{salter83,egger95}). 
The theories that better meet the observations 
interprete the 
NPS as the front shock of an evolved Supernova Remnant (SNR), 
whose distance should be $\sim 100\pm20$~pc
(inferred from starlight polarization, see \cite{bingham67}). 
In contrast, the present knowledge of the Fan Region
is much poorer.
A rough estimate of its distance can be derived from 
purely geometrical considerations: it has such an extent on the sky
that it must be located within (1-2)$\times 10^2$~pc
from the Sun not to have unrealistically large 
dimension (see also \cite{wolleben06}).
Another common and remarkable characteristic of the selected
areas is that the polarization vectors appear mostly aligned 
at the two higher frequencies of the Leiden surveys. 
Another region (patch 4) located at 
($70^\circ \leq l \leq 120^\circ$, $-45^\circ \leq b \leq -15^\circ$)
with a sky sampling in the average 
(its APS is then statistically relevant only for $\ell \lsim 50$)
but characterized by a rather low signal has been also 
considered by \cite{laportaburigana06}.

\begin{figure}[t]
\centering
\hskip -0.1cm
\includegraphics[width=5.cm,angle=90]{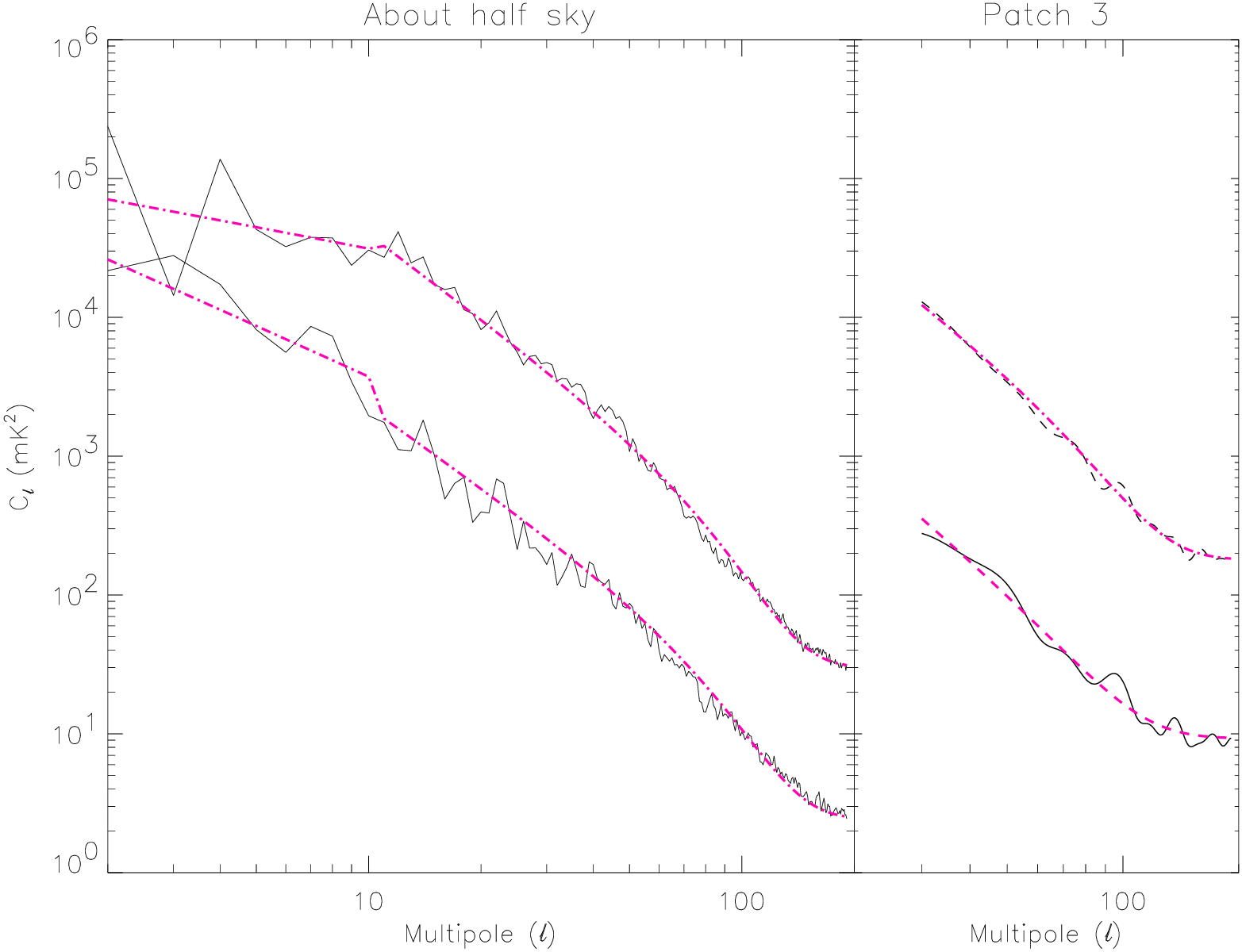}
\hskip 1.cm
\includegraphics[width=5.cm,angle=90]{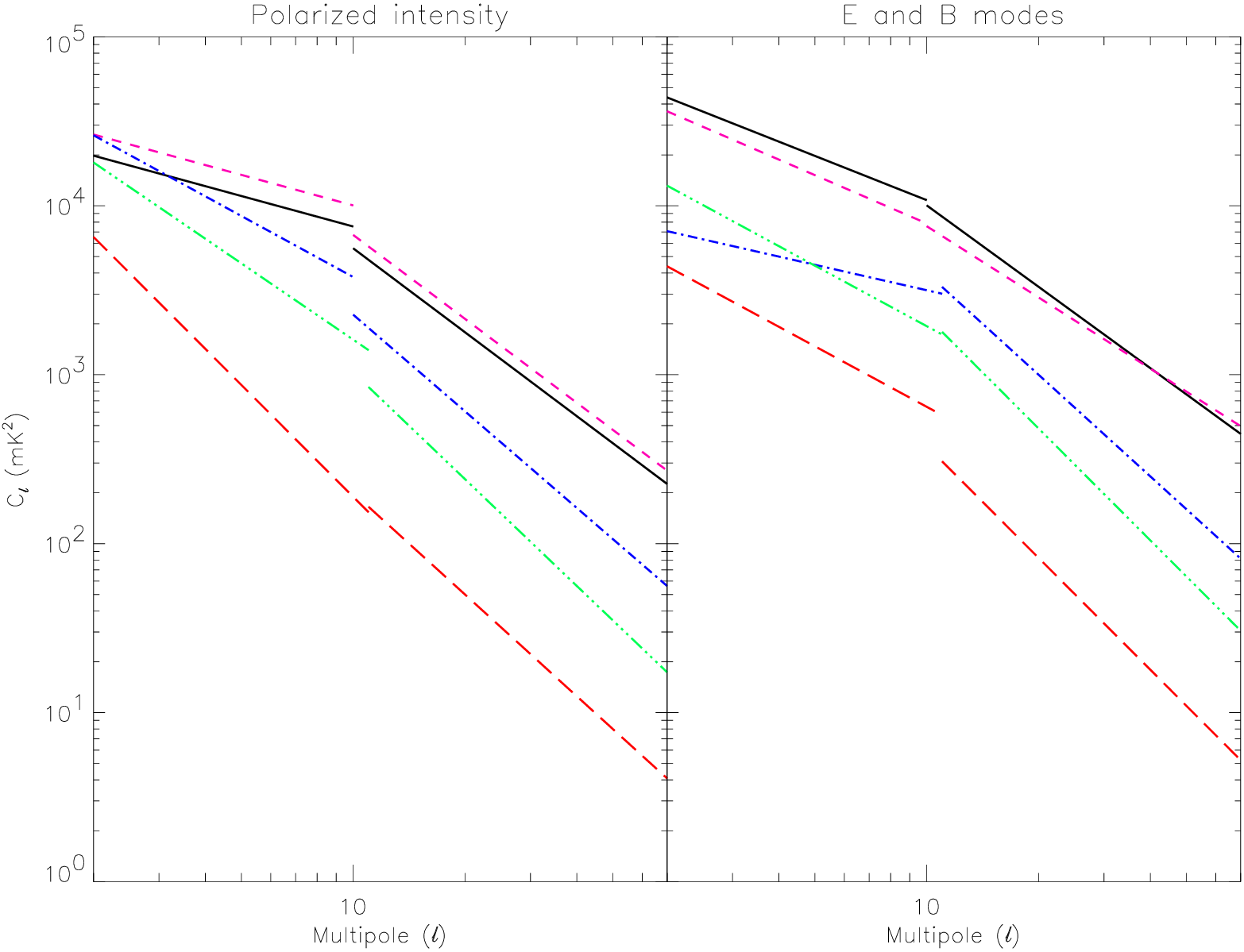}
\caption{Left panels: polarization APSs at 610~MHz (solid lines) for the survey 
full coverage (left panel) and patch 3 (right panel) together with the 
corresponding best fit curves (dot-dash lines). 
The lower curves in each panel are $C_{\ell}^{PI}$, 
while the other ones represent $C_{\ell}^{E;B}$
multiplied by 10.
Right panels: 
synchrotron component of the APS derived for the
survey full coverage for the polarized intensity 
(left panel) and for $C_{\ell}^{E;B}$ (right panel) 
at the various frequencies: 408 (solid line),
465 (dashes), 610 (dot-dash), 820 (three dot-dash), and 
1411~MHz (long dashes). Adapted from \cite{laportaburigana06}.}
\label{showfit610}
\end{figure}

\subsection{Angular power spectra}

In general, the $E$ and $B$ modes turned out to be extremely 
similar~\footnote{On the contrary, 
the $E$ and $B$ modes of the primordial CMB anisotropies
differ largely as they are induced by different mechanisms
 (e.g. \cite{seljak97}, \cite{Kosowsky99}.}, so that, 
instead of studying each of them
singularly, it can be more advantageous to consider their average
$C_{\ell}^{E;B}=(C_{\ell}^{E}+C_{\ell}^{B})/2$ (\cite{laportaburigana06}).
In some cases 
we could even consider a unique APS defined as 
$(C_{\ell}^{PI}+C_{\ell}^{E}+C_{\ell}^{E})/3$.
The maps essentially represent the Galactic polarized synchrotron emission, smoothed 
with the beam of the radiotelescope and contaminated by the noise.
As a consequence, their APSs can be fit as the sum of two components,
$C_{\ell} = C_{\ell}^{synch} W_{\ell} + C_{\ell}^{\rm N}$.
We exploit the power law approximation 
$C_{\ell}^{synch} \simeq \kappa \cdot \ell^{~\alpha} \, $ and
assume a symmetric, Gaussian beam, i.e.  
a window function
$W_{\ell}={\rm e}^{-(\sigma_{b} \ell)^2}$,
where 
$\sigma_b= { \theta_{HPBW} ({\rm rad}) / {\sqrt{8 {\rm ln} 2}} }$~.
Under the hypothesis of uncorrelated Gaussian random noise
(white noise),
$C_{\ell}^{\rm N} = C_{\ell}^{\rm WN}\sim$~const~.
The fit has been performed on the multipole range $[30,200]$ for 
the patches and $[2,200]$ for the full-coverage maps (\cite{laportaburigana06}).
In fact, 
the flattening occurring at higher multipoles helps
the recovering of the noise constant, in spite of the fact that
no reliable astrophysical information 
can be derived for ${\ell} > 100$ from the APSs even in the three better 
sampled regions (\cite{buriganalaporta02}).
For the patches, a single set of the parameters
$\kappa$, $\alpha$ allows to describe the APS in the entire
range of multipoles. In the case of the survey full 
coverage two sets of $\kappa$, $\alpha$ are needed, each of them
appropriate to a certain multipole range.
In fact, at all frequencies a change in the APS
slope occurs in correspondance of ${\ell} \sim 10$ for both  
$C_{\ell}^{PI}$ and $(C_{\ell}^{E}+C_{\ell}^{B})/2$.
As an example, Fig.~\ref{showfit610} displays the APS and 
the corresponding best fit curve at 610~MHz for the survey 
full coverage and one patch.
It shows also the recovered synchrotron term, $C_{\ell}^{synch}$,
for the survey.
Table~\ref{BestFitPar_patch} lists
the values of $\kappa$, $\alpha$ obtained 
for the patches.
In the multipole range 
$30 \lsim {\ell} \lsim 100$ 
a general trend appears: from the lowest to the highest frequency
the slope steepens  from $\sim -$(1-1.5) to $\sim -$(2-3), 
with a weak dependence on the considered sky region (\cite{laportaburigana06}).

\begin{table*}[!ht]
\begin{center}
\begin{tabular}{|c|c|c|c|c|c|c|c|c|c|c|}
\hline
&\multicolumn{2}{c|}{408~MHz}& \multicolumn{2}{c|}{465~MHz} 
       &\multicolumn{2}{c|}{610~MHz} &\multicolumn{2}{c|}{820~MHz} 
       &\multicolumn{2}{c|}{1411~MHz} \\   
& $\kappa$~(mK$^2$)& -$\alpha$ & $\kappa$~(mK$^2$) & -$\alpha$ & 
$\kappa$~(mK$^2$) 
& -$\alpha$ & 
       $\kappa$~(mK$^2$) & -$\alpha$ & $\kappa$~(mK$^2$) & -$\alpha$ \\
\hline
  1 & 
$9.80\cdot10^4$ & 0.99 & $6.00\cdot10^5$ & 1.50 &
            $2.21\cdot10^6$ & 2.05 & $3.08\cdot10^6$& 2.59 &
            $9.80\cdot10^5$& 2.90 \\
          & 
$1.00\cdot10^5$ & 1.10 & $6.00\cdot10^5$ & 1.67 & 
            $9.44\cdot10^6$ & 2.69 & $2.02\cdot10^6$ & 2.64 &
            $9.80\cdot10^5$  & 2.90\\
\hline
  2 & 
$1.26\cdot10^6$ & 1.81 &  $2.50\cdot10^6$ & 1.95 
            & $2.00\cdot10^7$& 2.66 & $2.58\cdot10^6$& 2.56  
            & $1.85\cdot10^6$& 3.13    \\
          & 
$1.26\cdot10^6$ & 1.10 & $1.00\cdot10^5$ & 1.32 & 
               $3.10\cdot10^5$ & 1.82 & $2.00\cdot10^6$ & 2.68 &
               $1.85\cdot10^6$  & 3.13  \\
\hline
  3 &  
$8.00\cdot10^4$  & 1.12 & $1.81\cdot10^7$ & 2.61 
           & $1.65\cdot10^6$ & 2.09 & $1.52\cdot10^6$& 2.56 
           & $1.70\cdot10^5$ & 2.45     \\
          & 
$1.20\cdot10^5$ & 1.61 & $8.00\cdot10^5$ & 1.32& 
              $9.00\cdot10^5$ & 2.28 & $3.53\cdot10^6$ & 2.93 &
              $2.70\cdot10^5$ & 2.10 \\
\hline
  4 &  
$2.00\cdot10^4$ & 0.91 & $2.60\cdot10^4$ & 1.08 
           & $3.10\cdot10^6$ & 2.69 & $3.36\cdot10^6$& 3.07 
           & $6.10\cdot10^5$ & 2.97     \\
          & 
$2.00\cdot10^4$ & 0.91 & $1.10\cdot10^5$ & 1.65 &
            $2.60\cdot10^5$ & 2.24 & $7.00\cdot10^4$ & 2.28 &
            $1.00\cdot10^6$ & 3.33 \\
\hline
\end{tabular}
\end{center}
\caption{Least-square best fit parameters $\kappa$ and $\alpha$
obtained in the APS analysis of the patches.
Relative errors on the best fit parameters are $\sim 10\%$. 
The considered multipole range is [30-200].
For each patch, the first (second) line refers to $C_{\ell}^{PI}$
($C_{\ell}^{E;B}$). Adapted from \cite{laportaburigana06}.}
\label{BestFitPar_patch}
\end{table*}

\section{Multifrequency APS analysis of Faraday depolarization effects}

The theoretical intrinsic behaviour of the synchrotron total intensity
emission predicts a power law dependence of the APS amplitude
on frequency:
$T^{synch}\propto \nu^{-\beta} \Rightarrow C_{\ell}^{synch}(\nu) \propto \nu^{-2\beta}\, ,$
where $\beta=\delta+2$. 
The values of $\delta$ deduced from radio observations change
with the considered sky position and range between 
$\sim 0.5$ and $\sim 1$ (see \cite{reich88,platania98}).
In principle, one would expect the synchrotron emission to be highly 
polarized, up to a maximum percentage of $\sim 75\%$ (\cite{ginzburg65}).
In the ideal case of constant degree of polarization
the above formulae would apply also to the
polarized component of the synchrotron emission, once the brightness 
temperature has been properly rescaled.
However, depolarization effects are relevant at radio frequencies 
and can mask a possible correlation 
between total and polarized intensity.
The current knowledge of $\beta$ for the Galactic diffuse 
polarized emission is very poor, but already indicates  
that due to depolarization phenomena the observed $\beta$ 
can be much lower than $\sim 2.7$ (e.g. for the NCP 
\cite{VinyajkinRazin2002} quote $\beta \sim 1.87$).
From the values of $C_{\ell=\tilde{\ell}}^{synch}(\nu)$ at 
various $\ell$, 
the APS amplitude is found to decrease with frequency,
but this frequency dependence is weaker than 
that observed in total intensity,
as expected in the
presence of a frequency dependent depolarization (\cite{laportaburigana06}).

An electromagnetic wave travelling through a magnetized plasma 
will undergo a change in the polarization status. 
Its polarization vector will be rotated by an angle
$\Delta\phi{\rm [rad]} = RM{\rm [rad/m}^2]\cdot \lambda^{2} {\rm [m}^2] \, ,$
where the rotation measure, $RM$, is the line of sight integral 
$RM [{\rm rad/m}^2]= 0.81 \int n_{e}[{\rm cm}^{-3}] 
\cdot B_{\Vert}[\mu{\rm G}] dl[{\rm pc}]  \, ,$
being $n_{e}$ the electron density and $B_{\Vert}$
the component of the magnetic field along the line of sight.
In the {\it slab model} (\cite{burn66}), 
differential rotation of the polarization angle
decreases the polarized intensity (Faraday depolarization)
and the observed and intrinsic polarization intensities are related
by 
$T^{p,obs}(\lambda)=T^{p,intr}(\lambda) \cdot 
|{\sin\Delta\phi}/{\Delta\phi}| \; .$
Being $\left({T^{p}}\right)^2 \propto C_{\ell}(\nu)$, the above 
formula reads:
$C_{\ell}(\nu)=C_{\ell}^{intr}(\nu) \cdot 
({\sin{\Delta\phi}}/{\Delta\phi})^2 \; ,$
where $C_{\ell}^{intr}(\nu) = 
\kappa \cdot{\ell}^{\alpha}\cdot\nu^{-2\beta} \, .$
At a given $\ell$, considering two frequencies
we have:
${ C_{\ell}(\nu_1) }/{ C_{\ell}(\nu_2) }=({\nu_1}/{\nu_2})^{-2\beta} 
\cdot ({\Delta\phi_2} / {\Delta\phi_1}) ^2 \cdot 
({\sin{\Delta\phi_1} / \sin{\Delta\phi_2}})^2 \; .$
Given $\beta$, the expression on the right side of
this equation can be 
computed as a function of $RM$
and the observed value of ${ C_{\ell}(\nu_1) }/{ C_{\ell}(\nu_2) }$
can provide informations about $RM$.

If the polarization angles vary within the beam area, then observations 
of the same sky region with different angular resolution will give 
an apparent change of the polarized
intensity with frequency because of {\it beamwidth depolarization};
on the contrary, bandwidth depolarization is negligible for these surveys
(see \cite{GardnerWhiteoak66,sokoloff}). 
Suppose $\theta_{HPBW,1}$ and 
$\theta_{HPBW,2} (< \theta_{HPBW,1})$
are the angular resolutions respectively at $\nu_1$ and $\nu_2$.
The effect of beamwidth depolarization on ${ C_{\ell}(\nu_1) }/{ C_{\ell}(\nu_2) }$
can be removed smoothing the map at the 
frequency $\nu_2$ to the (lower) angular resolution of the 
map at the frequency $\nu_1$ or, equivalently in practice, multiplying 
the observed APSs ratio, ${C_{\ell}(\nu_{1})}/{C_{\ell}(\nu_{2})}$ 
by ${\rm e}^{(\Delta\sigma_{b}\ell)^2}$, 
where $\Delta\sigma_{b}={\sigma_1}^2-{\sigma_2}^2$  
($\sigma_i=\theta_{HPBW,i}$~(rad)/${\sqrt{8 {\rm ln} 2}}\;$; $i=1,2$). 

Therefore, from 
$C_{\tilde{\ell}}^{synch}(820~{\rm MHz})/C_{\tilde{\ell}}^{synch}(1411~{\rm MHz})$
(case in which the oscillations of this theoretical ratio do not 
introduce the dramatic degeneration appearing at lower frequencies),
the following intervals of $RM$ are found to be compatible with the data: 
$\simeq$~9-17, 53-60, 75-87, 123-130~rad/m$^2$ together with two narrow 
intervals at $\simeq$~70 and 140~rad/m$^2$ (\cite{laportaburigana06}). 
A typical $RM$ value of $\simeq$~8~rad/m$^2$ has been reported 
by \cite{spoelstra84},
however this is likely a lower limit to the real values.
The possible values of the $RMs$
can be also constrained by the observed polarization degree, defined as 
$\Pi=T^{p}/T\; .$
The mean polarization degree
$\overline{\Pi}$ in the portion 
of the sky covered by the Leiden survey at 1411 MHz
can be estimated exploiting the total intensity northern sky map
at 1.4~GHz (\cite{reich82,reich86}):
$\overline{\Pi}^{obs}\sim 17, 27, 28, 25, 12\%$ for the full 
coverage and for the patch 1, 2, 3, and 4, respectively. 
Given the relation between the intrinsic and
observed brightness polarized temperature in the slab model
approximation, we have:
$\Pi^{obs}=\Pi^{intr} \cdot|\sin\Delta\phi/\Delta\phi|\; .$
The maximum possible value of $\Pi^{obs}$  
(derived assuming $\Pi^{intr}=0.75$) can be computed
as function of the $RM$
and compared with 
the mean polarization degree actually observed 
in the considered cases\footnote{Note that
the observed mean polarization degree might be due to the combination
of Faraday depolarization and beam depolarization. In case we were
able to correct for beam depolarization effects
this analysis will lead to 
smaller values of RMs.
Therefore the results of our analysis are rather conservative 
and provide upper limits to the RMs compatible 
with observations in the considered patches. }. 
The observed polarization degree 
sets the upper limit $RM \lsim$~50~rad/m$^2$,
consistent with the lowest of the $RM$ intervals previously 
identified (\cite{laportaburigana06}).

\begin{figure}[t]
   \centering
   \begin{tabular}{ccc}
   \includegraphics[width=3.1cm,angle=90,clip=]{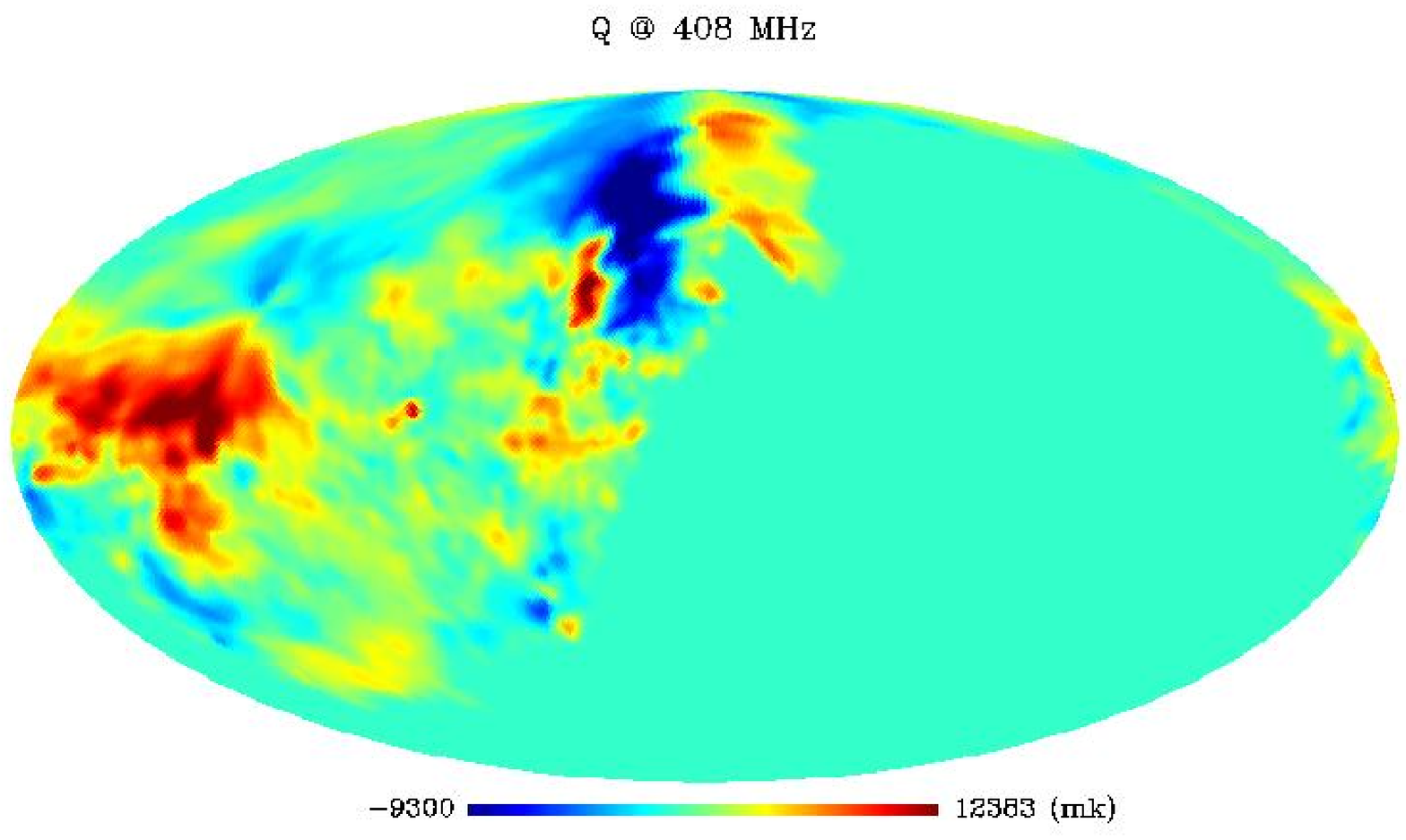}
   \includegraphics[width=3.1cm,angle=90,clip=]{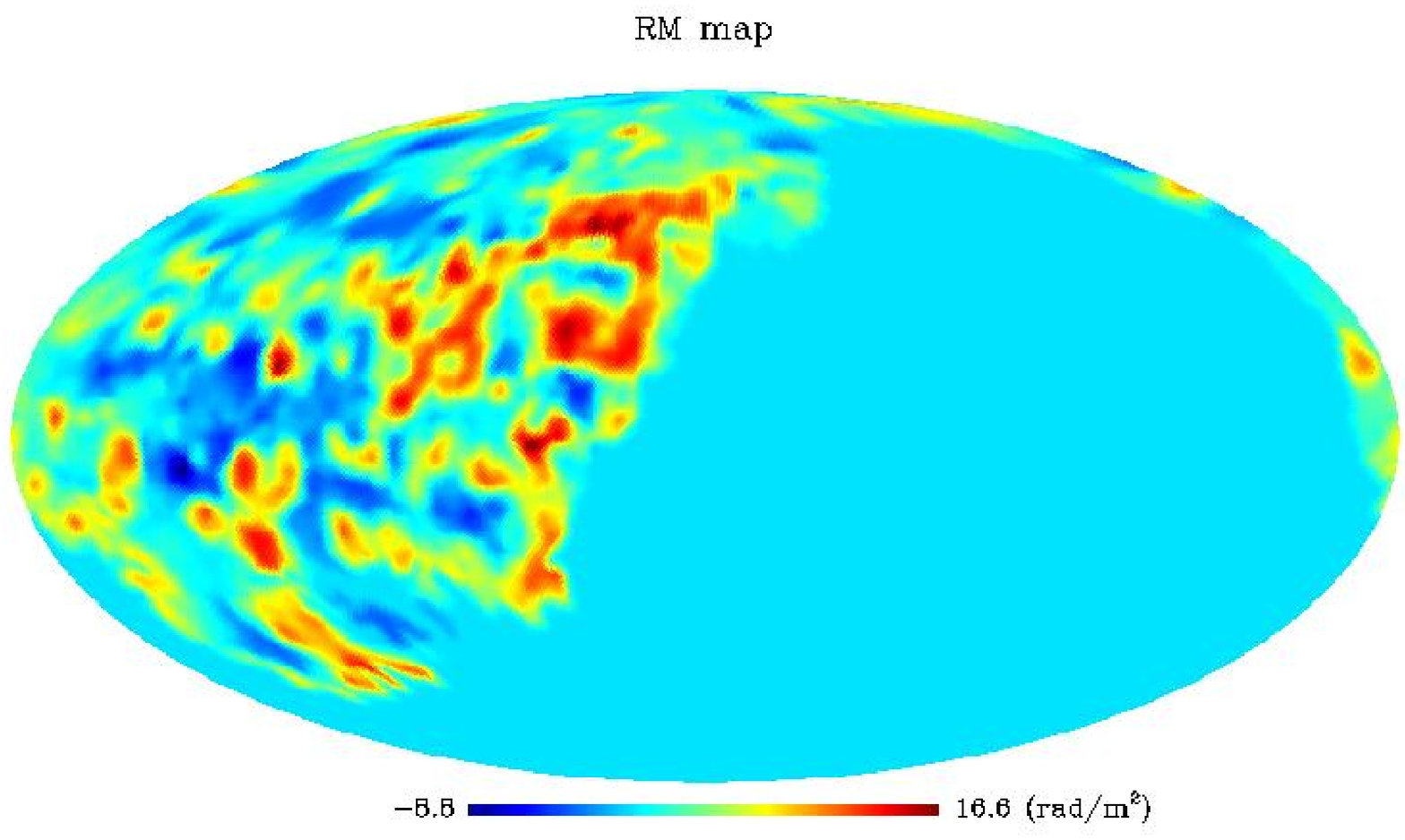}
   \includegraphics[width=3.1cm,angle=90,clip=]{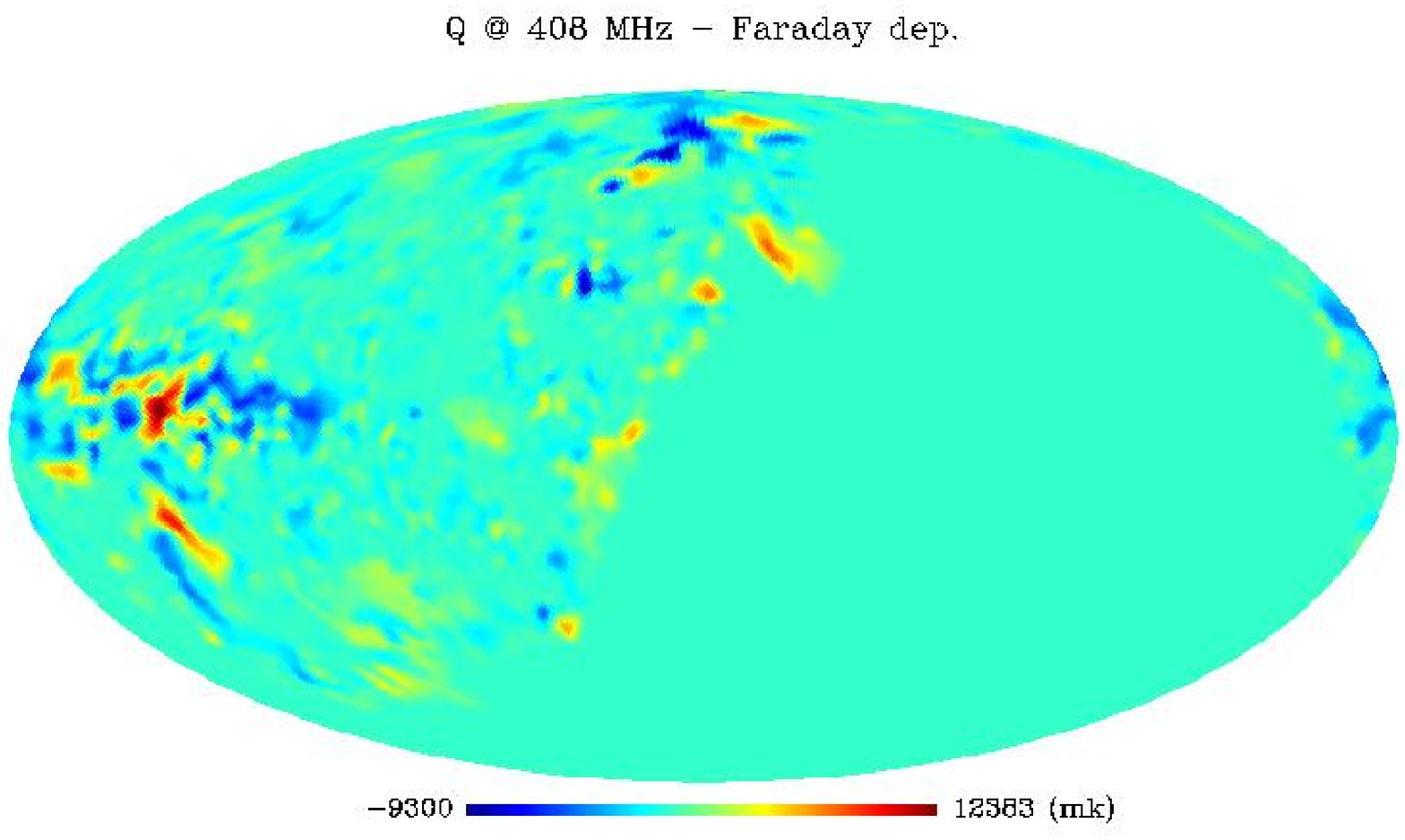} 
   \end{tabular}
   \caption{Maps of the $Q$ Stokes parameter at 408~MHz before (left panel) and 
    after (right panels) the transformation for Faraday depolarization
    through the $RM$ map (middle panel) used in the simulation.
Similar results are found for the $U$ Stokes parameter.
    All maps have been smoothed to $2.3^{\circ}$. 
Adapted from \cite{laportaburigana06}.
}
   \label{DepMap408}
 \end{figure}

As already shown, the synchrotron 
emission APS steepens with increasing frequency.
One possible explanation of this
behaviour
again relies on depolarization arguments:
electron density and/or magnetic field fluctuations (in strength 
and/or direction) might redistribute the 
synchrotron emission power from larger to smaller angular 
scales, creating fake structures,
the effect being less important at increasing frequency.
Sticking to the ISM {\em slab model}, 
the Faraday depolarization effects on the APS
at 408~MHz and 1420~MHz can be simulated as follows. 
The 1420~MHz polarization map (\cite{wolleben06}) 
is assumed to
represent the intrinsic Galactic synchrotron emission and is 
scaled down to 408~MHz
with a spectral index of $-2.75$.
The 
intrinsic
polarization angle, 
as mapped by the polarization angle map at 1420~MHz,
should be the same at the two frequencies. 
An $RM$ map constructed exploiting 
the catalog produced by T.A.T.~Spoelstra (private communication), 
consisting of $\sim 1000$~~$RM$ data,
has been adopted 
to transform the intrinsic $\phi$ and $PI$ for  
Faraday rotation and differential Faraday depolarization:
$\phi_{output} \to \phi_{input} + \Delta\phi$ and
$PI_{output} \to PI_{input} \cdot |sin(\Delta\phi)/\Delta\phi|$.
At 1420~MHz the input and output maps are very similar.
At 408~MHz 
the large scale structure appears significantly 
depressed in the depolarized map, whereas the small scale 
features become much more relevant.
The results of this toy-simulation are shown 
in  Fig.~\ref{DepMap408} for the Stokes parameter $Q$.
The comparison between the APSs of the input and output maps
quantifies this behaviour (\cite{laportaburigana06}):
at 1420~MHz the APSs are almost identical
while at 408~MHz the depolarized map APSs significantly flatten,
in agreement with the 
derived APS flattening with decreasing frequency
reported in Table~\ref{BestFitPar_patch}.

\section{Analysis of the DRAO 1.4 GHz polarization survey}

A new linear polarization survey
of the northern celestial
hemisphere at 1.42 GHz with
an angular resolution
$\theta_{HPBW} \simeq 36'$
has been recently completed
using the DRAO 26~m telescope
(\cite{wolleben06}).
The survey
has a spacing, $\theta_s$, of $15'$ in right ascension
and from $0.25^\circ$ to $2.5^\circ$
in declination,
which requires interpolation to construct equidistant cylindrical 
({\tt ECP}) maps with $\theta_{pixel} = 15'$ (\cite{wollebenPhD}). 
The final maps have an angular resolution of about 
$36'$ and an rms-noise of about 12 mK, which is
unique so far
in terms of sensitivity and coverage.
These polarization data have been used by \cite{laportaetal06} in combination with the 
Stockert total intensity map at 1.42 GHz
(\cite{reich82,reich86}),
having the same angular resolution, similar sensitivity, and $\theta_s = 15'$.

\subsection{Selected areas and results}

Being interested in the diffuse Galactic synchrotron emission, 
\cite{Buriganaetal2006} subtracted discrete sources (DSs) from the total 
intensity map (only very few sources are visible 
in polarization out of the plane) 
performing a 2-dimensional Gaussian fitting (NOD2-software library, \cite{haslam74}).
We are confident that at $|b| \gsim 45^{\circ}$ DSs with peak flux densities 
exceeding $\sim 1$ Jy have been removed that way 
(see \cite{Buriganaetal2006} for a map of the subtracted DSs).

\begin{figure}[t]
\hskip -0.2cm
   \begin{tabular}{ccc}
   \includegraphics[width=2.cm,angle=0,clip=]{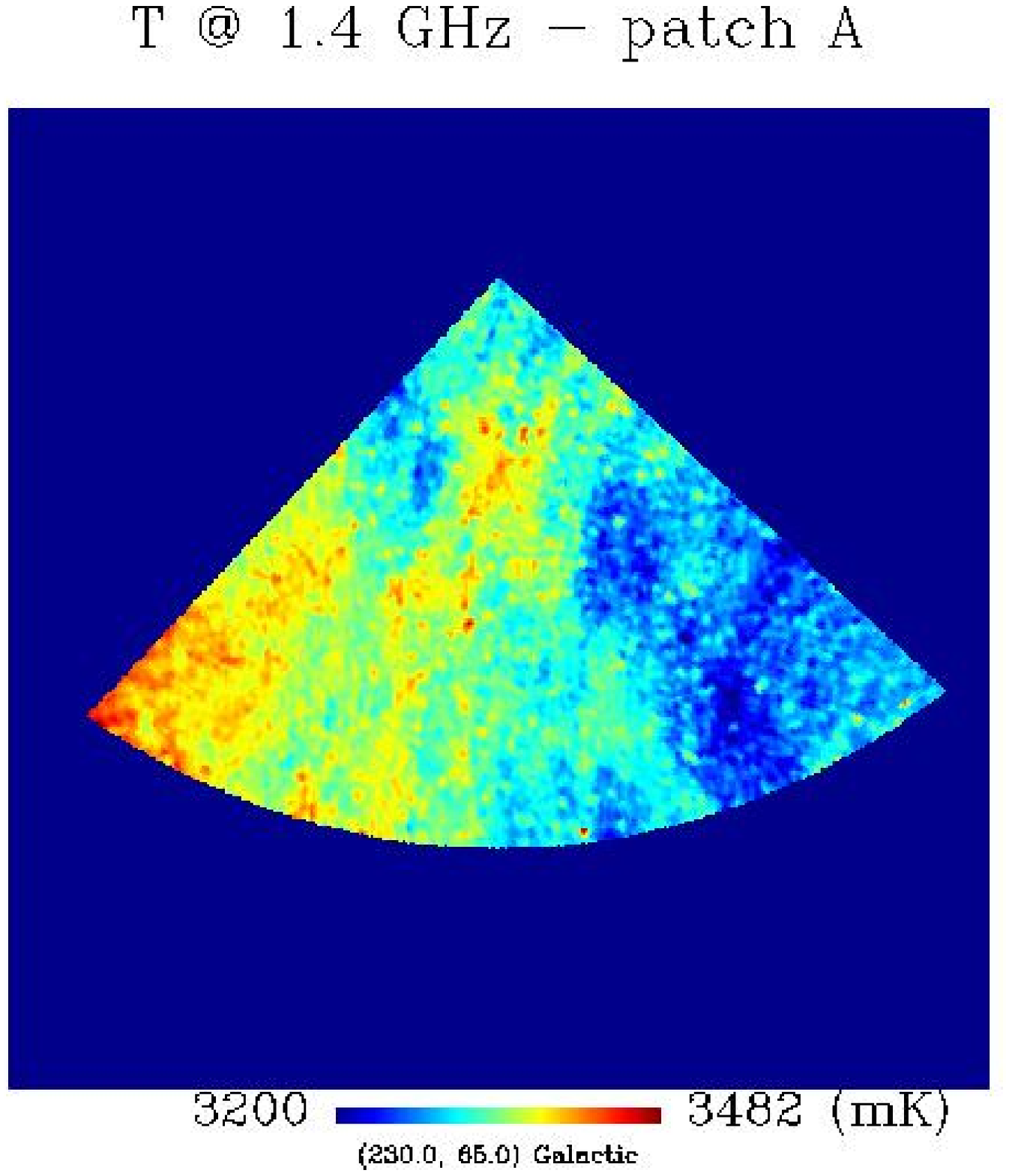}&
   \includegraphics[width=2.cm,angle=0,clip=]{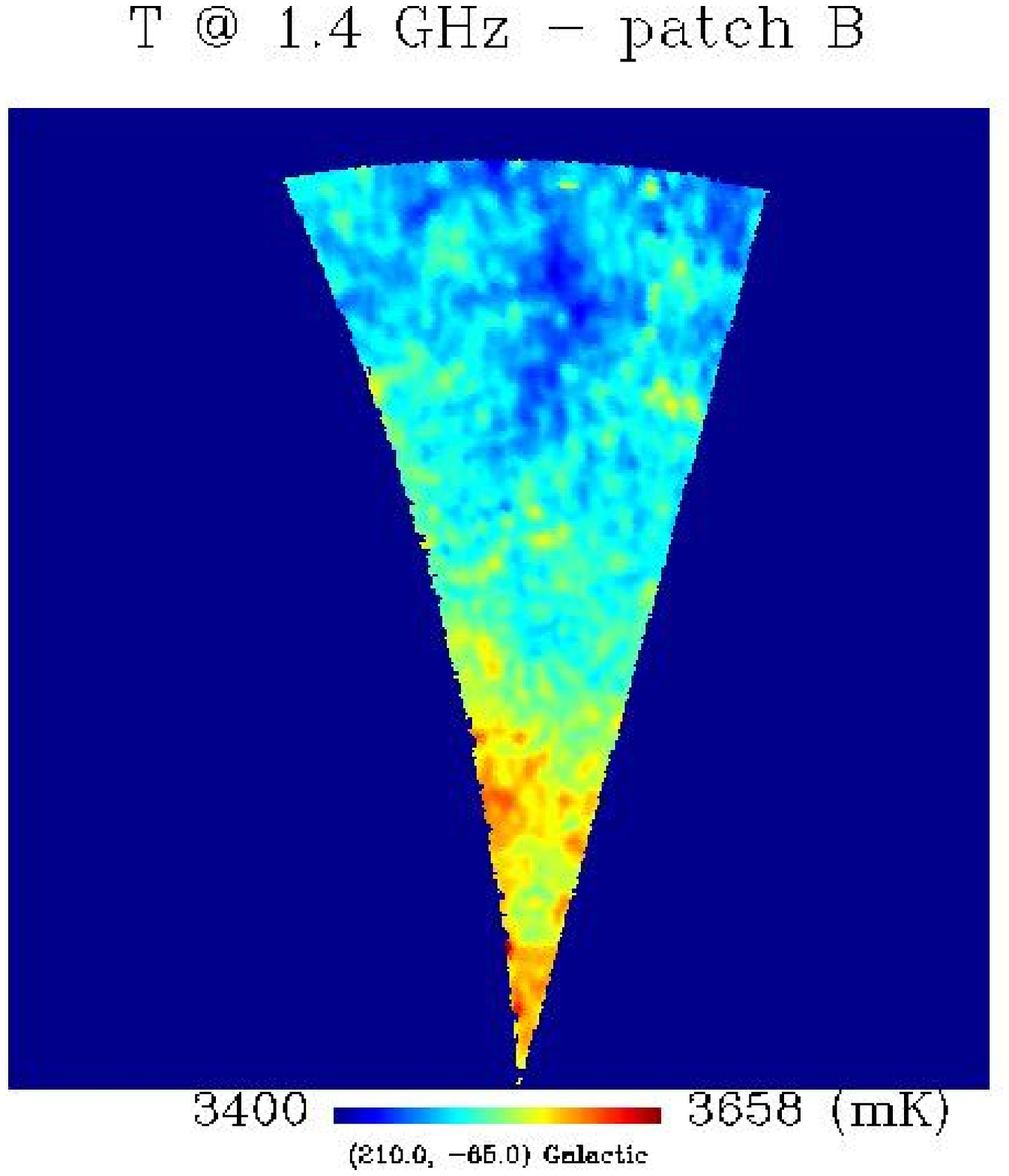}&
   \includegraphics[width=2.cm,angle=0,clip=]{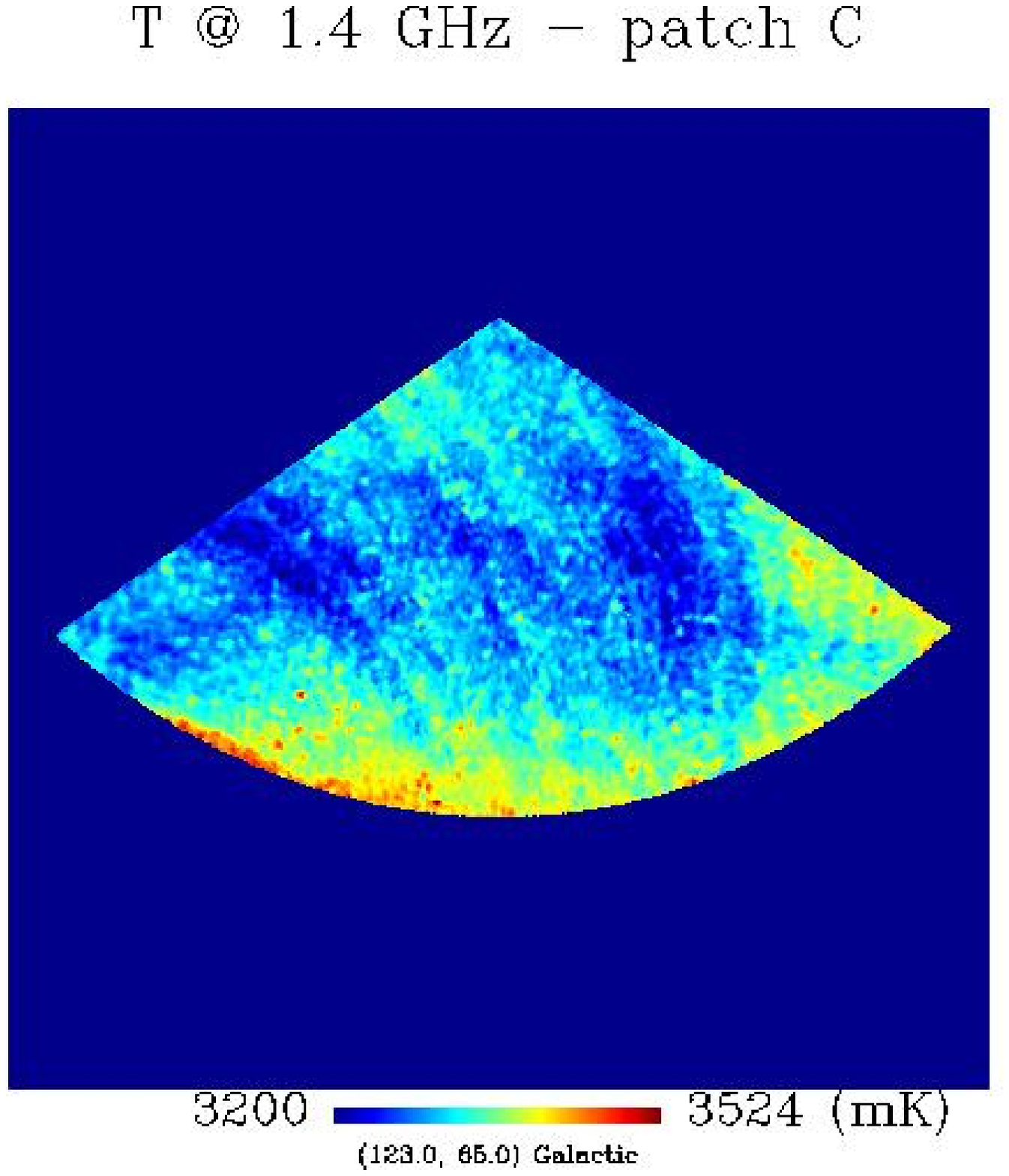}\\
   \includegraphics[width=2.cm,angle=0,clip=]{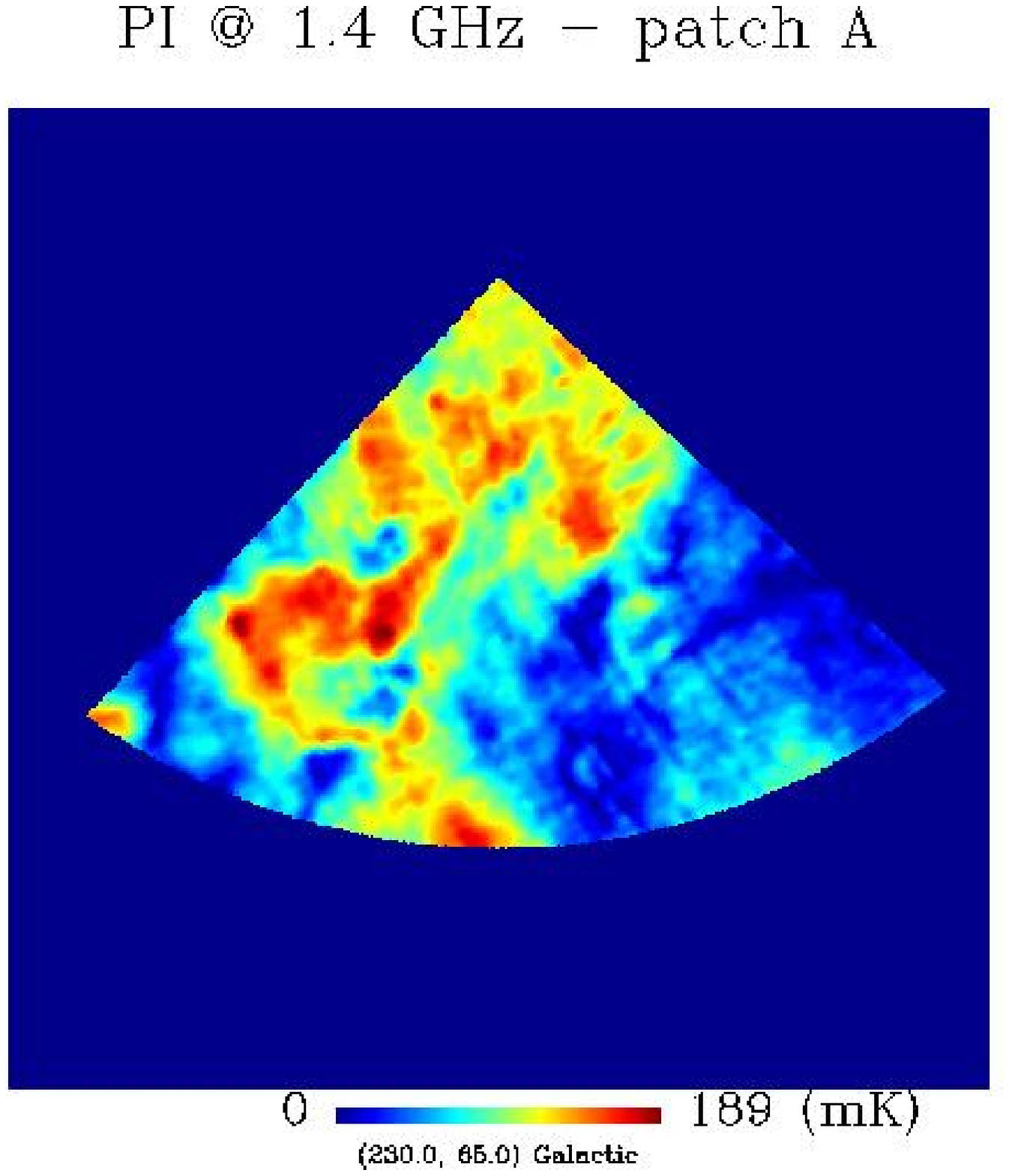}&
   \includegraphics[width=2.cm,angle=0,clip=]{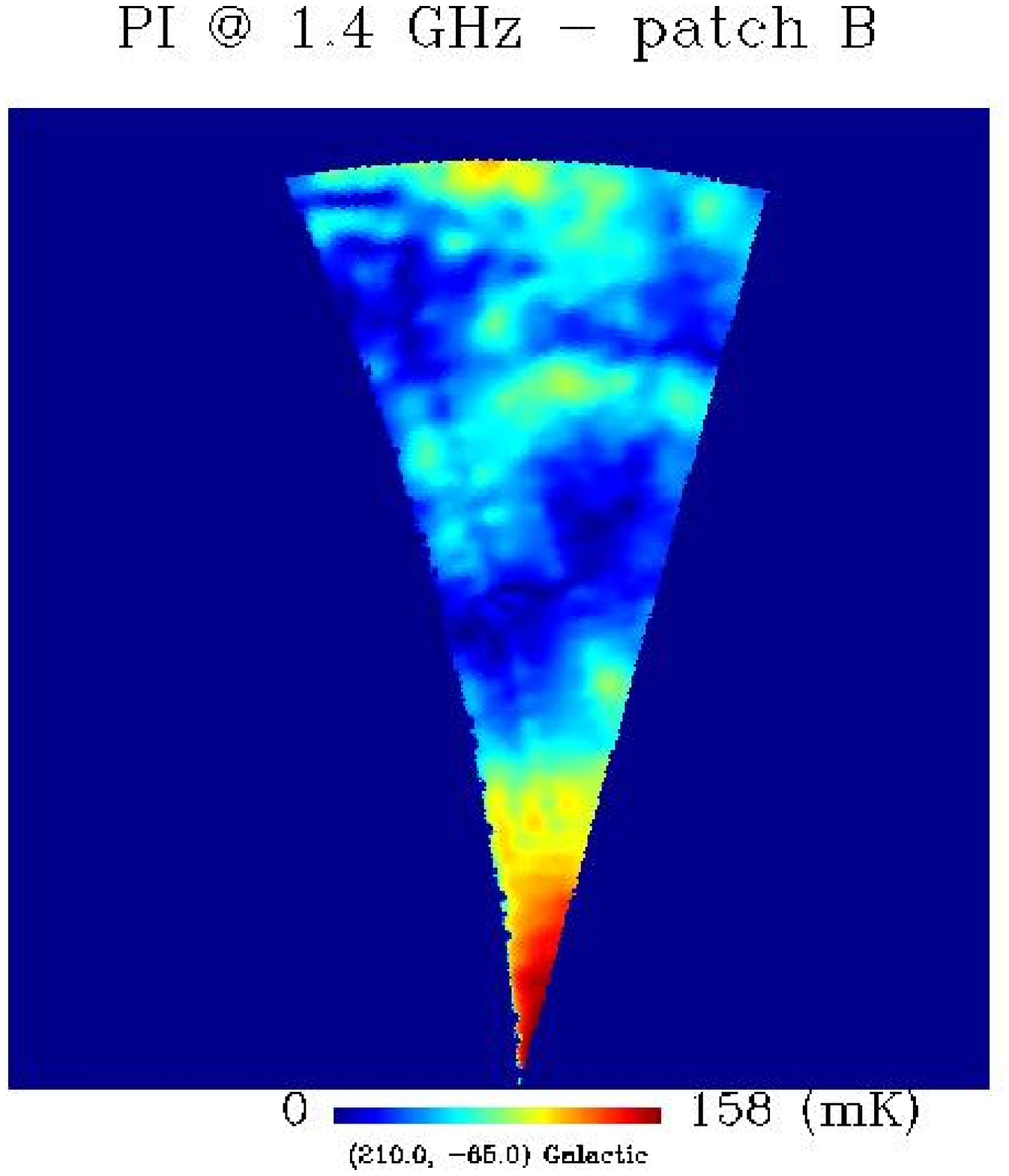}&
   \includegraphics[width=2.cm,angle=0,clip=]{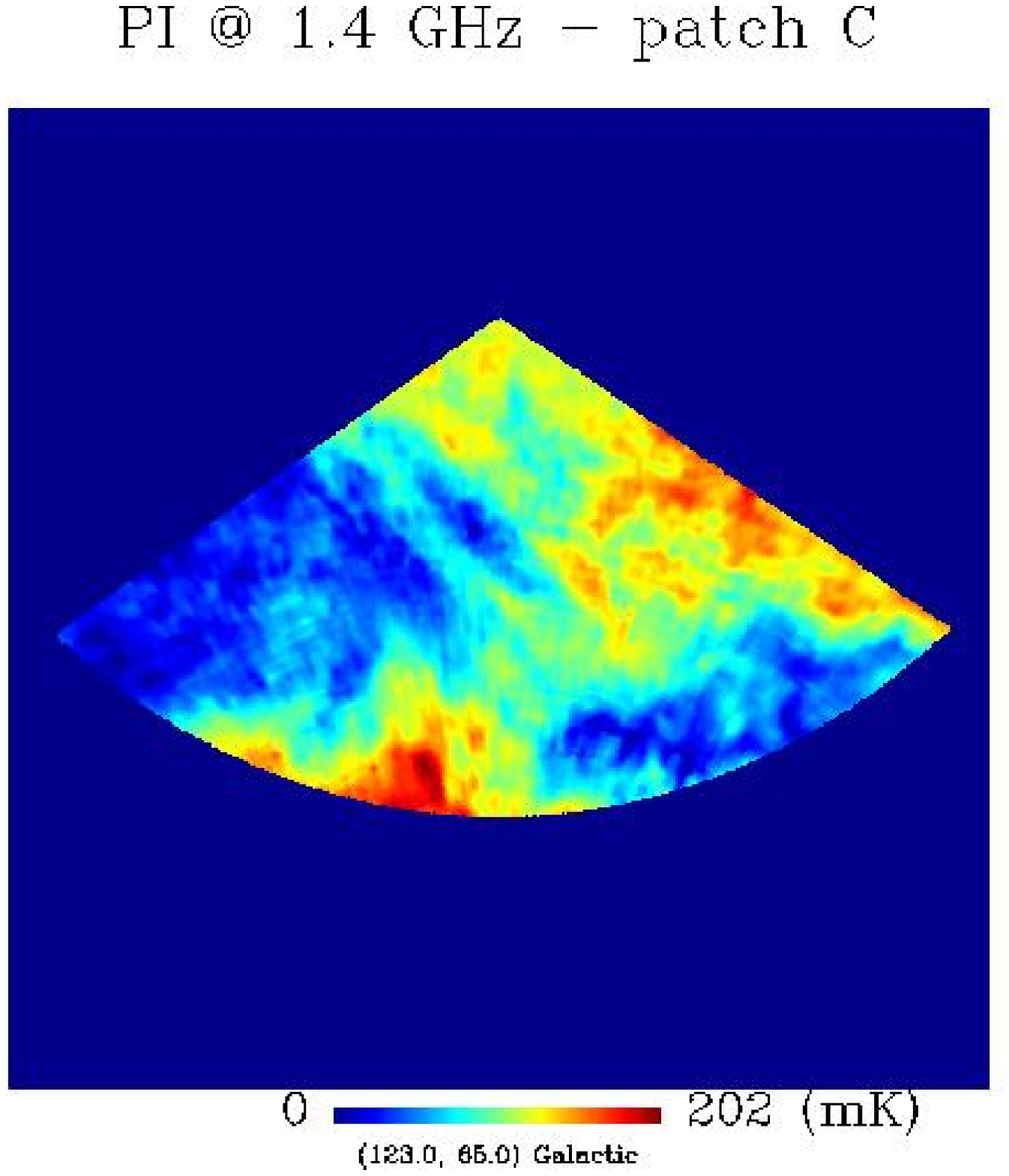}\\
   \end{tabular}
   \begin{tabular}{c}
\hskip 0.5cm
   \includegraphics[width=4.5cm,height=7cm,angle=90]{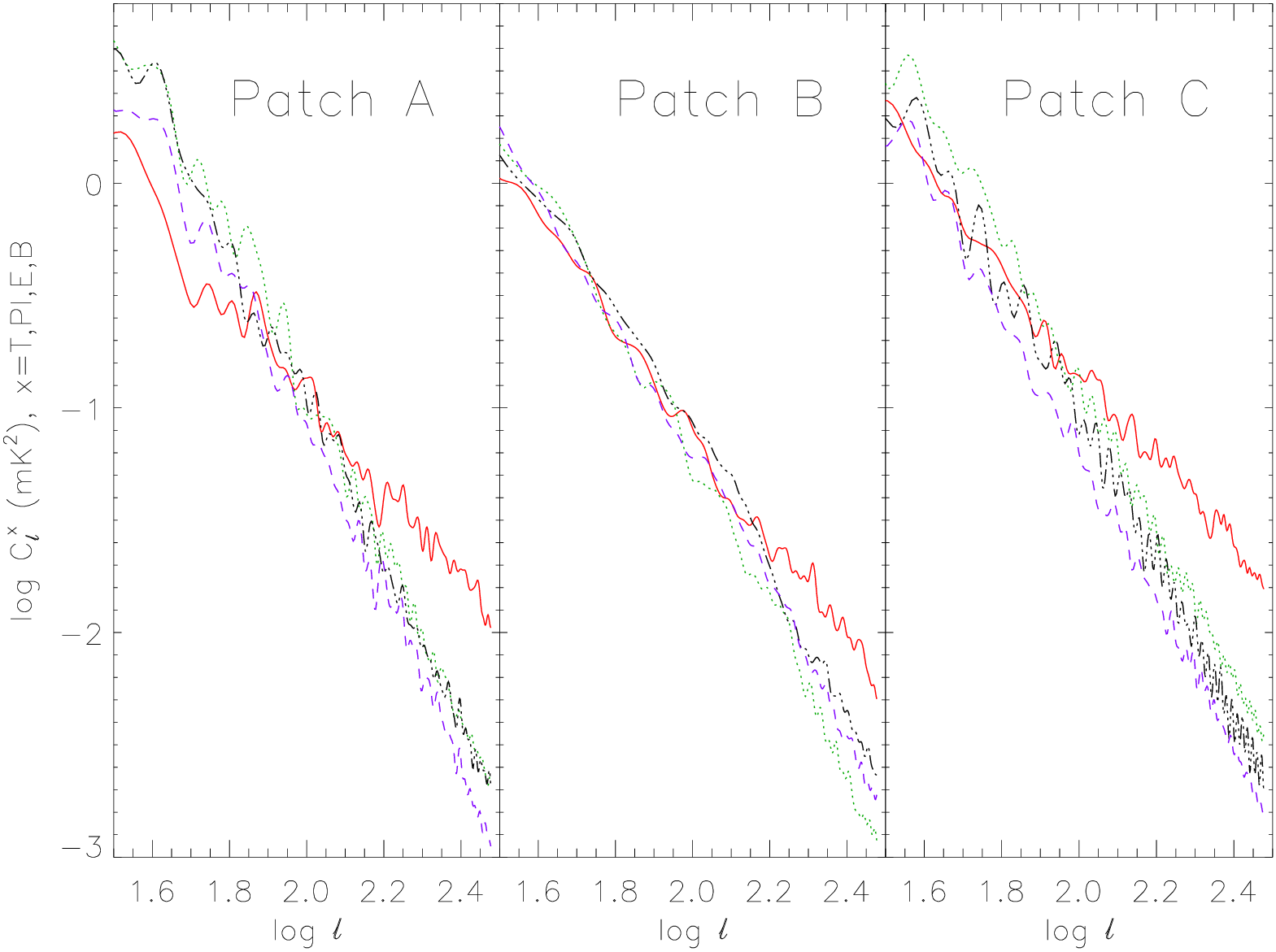}
   \end{tabular}
   \caption{Left small panels: gnomonic projection of the considered cold patches 
in total intensity (top panels) and polarization (bottom panels).
Right panels: APSs 
of   the selected cold regions 
  (solid lines (red lines) $\to C_{\ell}^{T}$; dashes (blue lines) $\to C_{\ell}^{PI}$;
  dots (green lines) $\to C_{\ell}^{E}$; three dot-dashes (black lines) $\to C_{\ell}^{B}$).
Adapted from \cite{laportaetal06}.}
   \label{patches}
 \end{figure}

We then derived the APSs
in both temperature and polarization 
for the whole DRAO survey coverage and the three following sky areas 
(see Fig.~\ref{patches})
characterized by a relatively weak polarized intensity:
patch A - $180^{\circ} \le l \le 276^{\circ}$, $b \ge  45^{\circ}$;
patch B - $193^{\circ} \le l \le 228^{\circ}$, $b \le -45^{\circ}$;
patch C - $ 65^{\circ} \le l \le 180^{\circ}$, $b \ge  45^{\circ}$.
In all patches the polarization APSs 
are rather similar (\cite{laportaetal06}).
This fact might indicate that depolarization due 
to differential Faraday rotation should not be relevant in such 
regions of the sky with respect to the investigated
angular scales (\cite{laportaburigana06}). 
In fact, rotation measure (RM) maps (\cite{johnston,dineen}) 
obtained interpolating RM data of extragalactic 
sources show very low values in correspondence of such areas.
However, the degree of polarization
 is on average a few percent, 
well below the maximum theoretical value of $\sim 75\%$.
The reason for the low fractional polarization is not clear. 
One possibility
is that depolarization other than 
differential is present (e.g.~\cite{sokoloff}).

The APSs 
(namely $C_{\ell}^{T}$, $C_{\ell}^{PI}$ and 
$C_{\ell}^{E;B}$)
can be modelled as 
$C_{\ell} = (C_{\ell}^{synch}+ c^{src})\cdot W_{\ell} + c^{WN} \,$ (\cite{laportaetal06}). 
The power law approximation 
$C_{\ell}^{synch} \simeq \kappa \cdot \ell^{~\alpha} \, $ 
has been exploited. 
The contribution of non-subtracted DSs has been modelled with a constant
term according to the formalism of 
Poisson fluctuations from extragalactic point sources
(\cite{toffolatti98}).
From extragalactic source counts at 1.4~GHz 
(see e.g. \cite{prandoni01}) we estimate
$c^{src} \sim 0.09$~mK$^2$
for the total intensity case. 
We assumed again a symmetric Gaussian beam window function.
A first guess estimate of the noise term $c^{WN}$ 
is provided by $C_{\ell}^{WN} \sim 4\pi \cdot \sigma_{pix}/N_{pix}$, 
where $N_{pix}$ is the number of pixels in the {\tt HEALPix} map and  
$\sigma_{pix}=\sigma_{pix,{\tt HEALPix}} \sim N^{-1/2} \cdot \sigma_{pix,{\tt ECP}}$
with $N$ number of the {\tt ECP} pixels corresponding to each
{\tt HEALPix} pixel at a given latitude.

The resulting best fit parameters for 
one patch are listed in Table~\ref{BFtab}. 
\begin{table}
\begin{center}
\begin{tabular}{|c|c|c|c|c|}
\hline
\multicolumn{2}{|c|}{Coverage} & \multicolumn{3}{|c|}{Best Fit parameters}\\
\hline
\multicolumn{2}{|c|}{     } &  ${\rm log}_{10}\kappa$ (mK$^2$) & $\alpha$ & $c^{src}$ (mK$^2$) \\
\hline
C & $C_{\ell}^{T}$  & $4.339^{-0.297}_{+0.284}$ 
        & $-2.64^{+0.03}_{-0.01}$ 
        & $0.047^{-0.030}_{+0.013}$\\ 
        & $C_{\ell}^{PI}$ & $4.748^{-0.350}_{+0.650}$ 
	& $-2.94^{+0.20}_{-0.36}$ 
	& $0.004^{-0.004}_{+0.004}$\\ 
        & $C_{\ell}^{E;B}$ & $4.954^{-0.301}_{+0.444}$ 
	& $-2.94^{+0.20}_{-0.26}$ 
	& $0.005^{-0.005}_{+0.009}$\\    
\hline
\end{tabular}
\end{center}
\caption{
         Least-square best fit parameters describing the total intensity 
	 and polarization APSs 
for the patch C in the range $30 \le \ell \le 300$.
	 The errors are given so that the upper (resp. lower) values correspond to the 
	 best fit parameters set with the flattest (resp. steepest) slope.
 Adapted from \cite{laportaetal06}.         }
\label{BFtab}
\end{table}
The parameter range for the Galactic synchrotron emission APS 
is $\alpha \sim [-2.71,-2.60]$ for the slope
and ${\rm log}_{10}\kappa \sim [4.067,4.339]$ ($\kappa$ in mK$^{2}$)
for the amplitude. 
The DS contribution is in the range $c_{src} \sim [0.023,0.047]$~mK$^{2}$,
which is consistent, within the errors, with the above source count 
estimate~\footnote{
The bulk of discrepancy of a factor
$\sim 2$ between the value of
$c^{src}$ recovered by our fit and that predicted
from best fit number counts can be in fact produced
by the survey sky sampling ($\theta_{s}$) of
$15'$.}.
For polarization, the derived slopes are generally steeper,
varying in the interval $\alpha \sim [-3.02,-2.62]$
for $C_{\ell}^{PI}$ and $\alpha \sim [-3.05,-2.55]$ 
for $C_{\ell}^{E;B}$ (\cite{laportaetal06}). 
The DS contribution is on average much lower
than in temperature,
being clearly compatible with zero~\footnote{The best fit results may suggest
a polarization degree (obtained considering
also the contribution of the subtracted DSs, $\sim 0.05 - 0.2$~mK$^2$,
to the temperature APS) considerably higher
than $\sim 2$\%,
the value found for NVSS extragalactic (mainly steep spectrum)
sources  (\cite{mesa02,tucci04}).
It may imply
 a presence (or a combination) of spurious instrumental polarization
at very small scales,
of a significant contribution from highly polarized Galactic sources
(\cite{manchester98})
non-subtracted in the maps, or of a flattening
of the diffuse synchrotron polarized emission APS at
$\ell \gsim 200-250$
with respect to the behaviour
at $\ell \lsim 200-250$, as that found
in higher resolution analyses on smaller sky areas
(\cite{bacci01,carretti06}).}.
The possibility to apply a source subtraction exploiting 
dedicated point source templates (\cite{Massardietal2006})
produced from existing catalogs is under investigation.

\section{Local analysis of the all-sky surveys}

\begin{figure}[t]
   \centering
   \begin{tabular}{cc}
   \includegraphics[width=4.cm,angle=90,clip=]{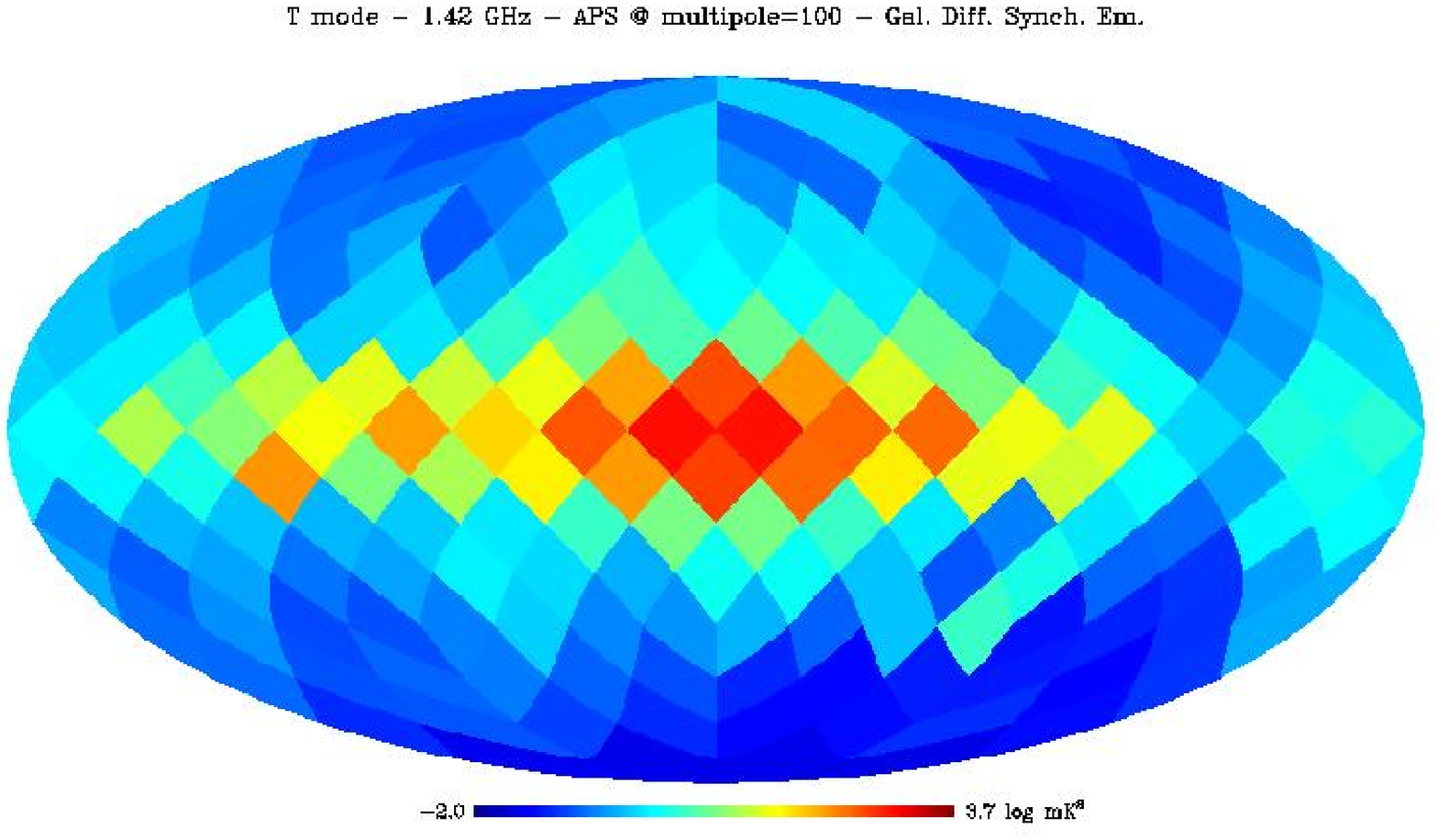}
   \includegraphics[width=4.cm,angle=90,clip=]{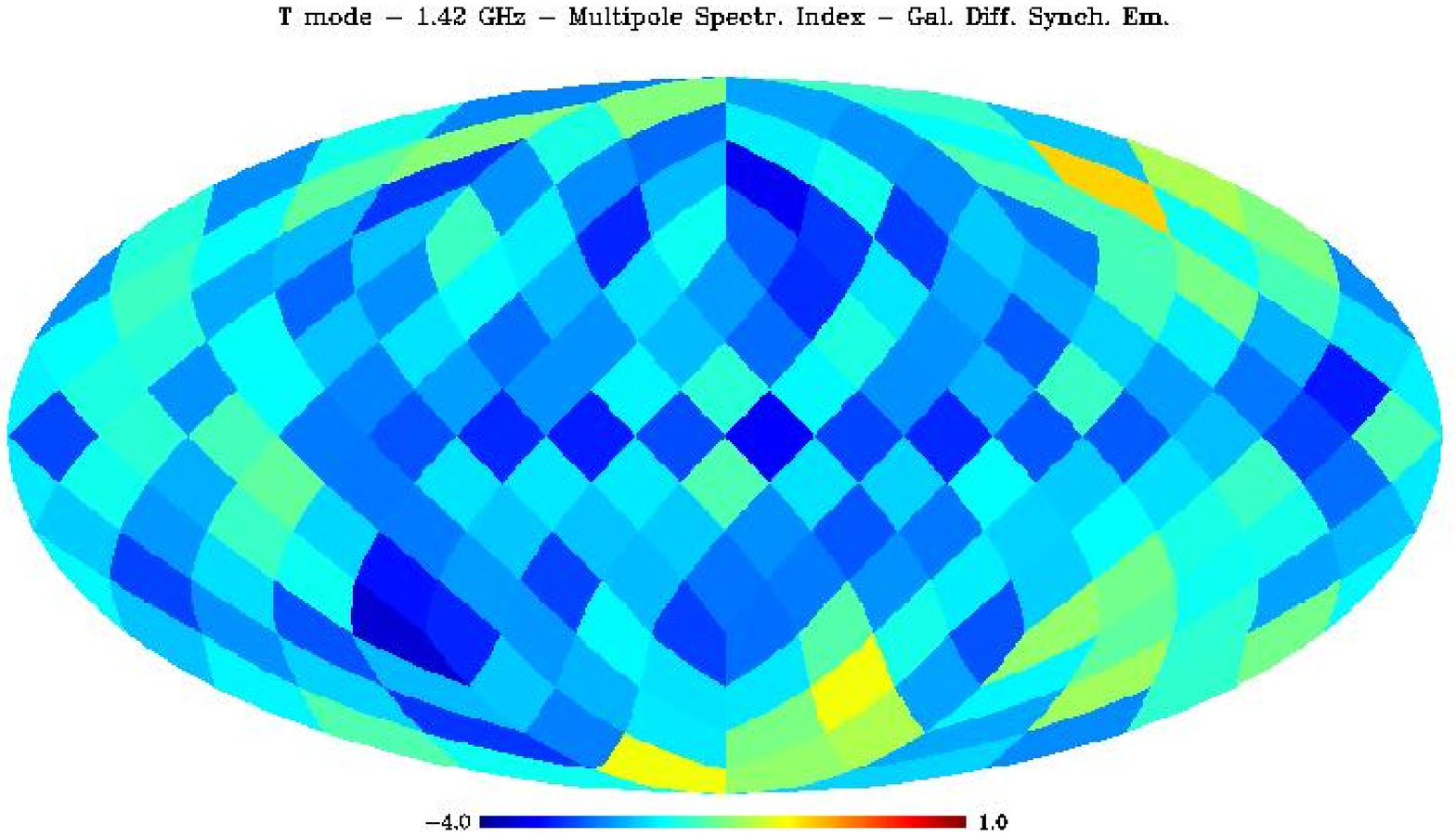}\\
   \includegraphics[width=4.cm,angle=90,clip=]{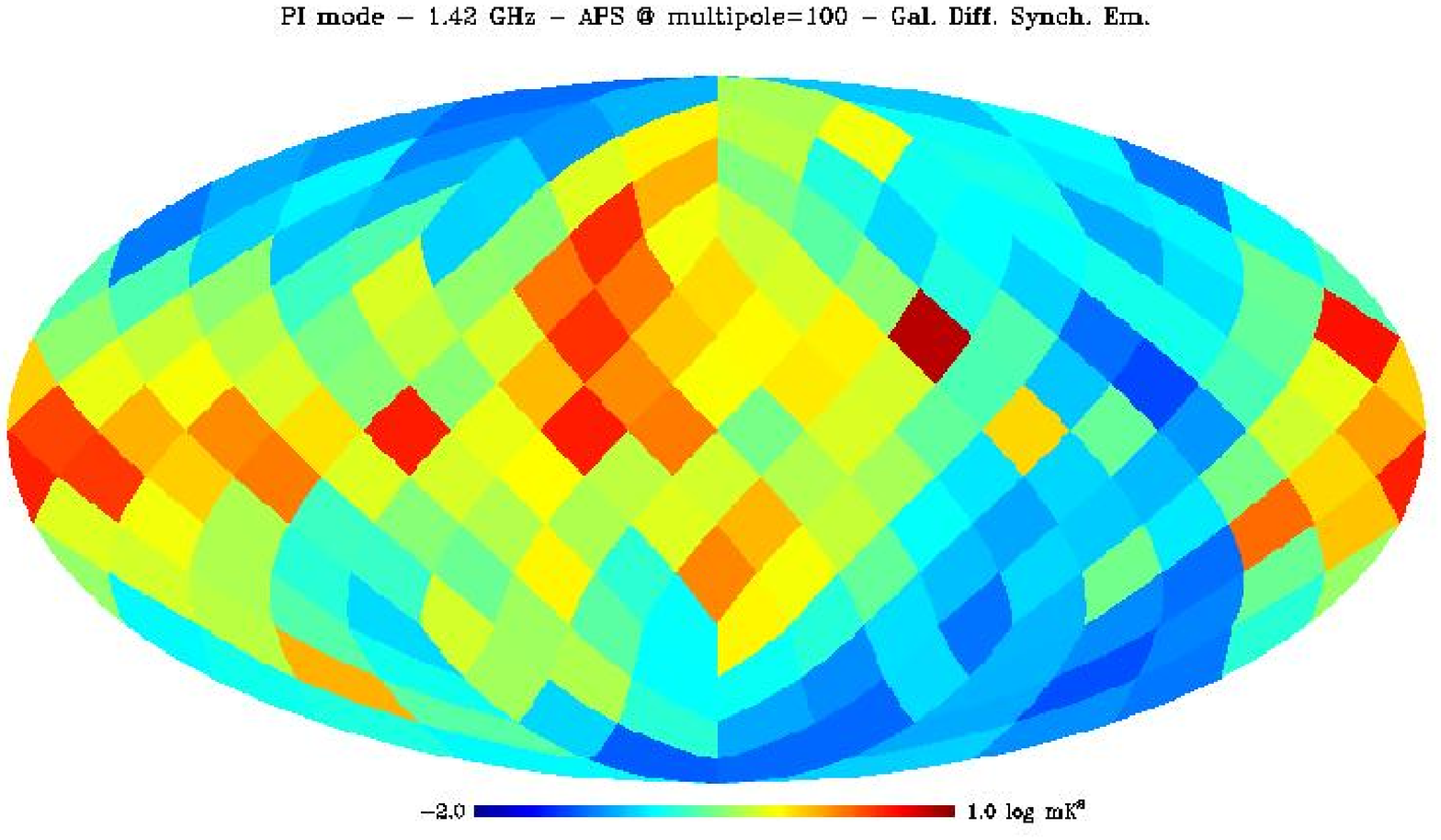}
   \includegraphics[width=4.cm,angle=90,clip=]{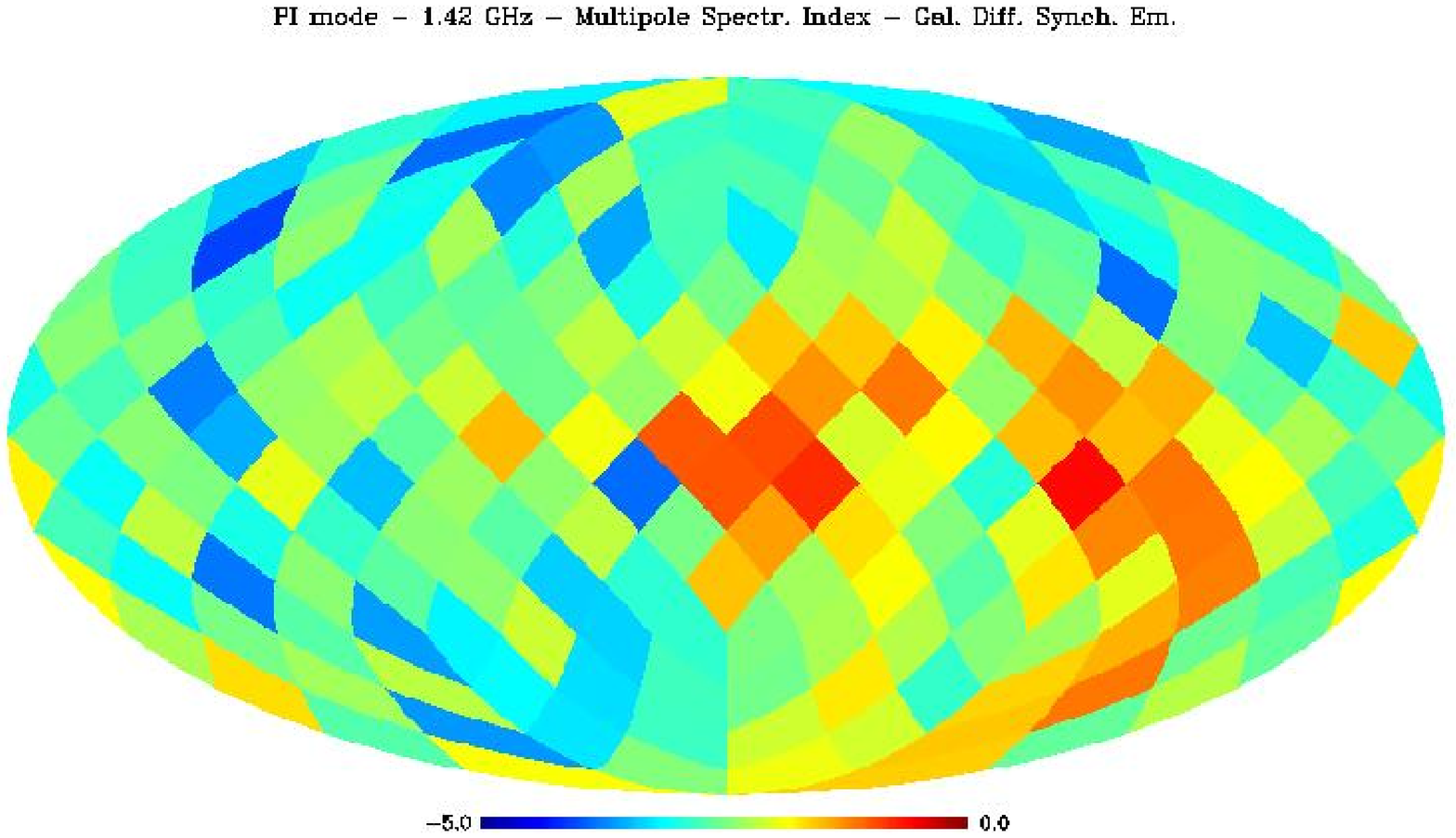}
   \end{tabular}
   \caption{Results of the local APS analysis for the total and polarization intensity 
maps. The results for the $E$ and $B$ modes are very similar to those found for $PI$.}
   \label{analysis_allsky}
 \end{figure}

In the recent years a complete coverage of the radio sky at 1420~MHz,
both in total intensity and in polarization intensity, has been
achieved. It derives from the combination
of the Stockert total intensity survey (\cite{reich82,reich86}),
of two DRAO polarization survey previously discussed, 
and of the Villa Elisa total intensity (\cite{testori01})
and polarization (\cite{testori03})
survey which covers the southern celestial
hemisphere with the same resolution, similar sensitivity, 
and with a uniform sampling ($\theta_s = 15'$).
A first APS analysis of the preliminary version of these all-sky surveys 
for representative Galactic cuts has been presented in \cite{Buriganaetal2006}.
Great efforts
have been recently dedicated to the intercalibration 
of the various sky areas and
further improvements are forseen.  
The present analysis is based on an updated version of the all-sky polarization
survey. 
As evident, diffuse synchrotron 
correlation properties depend on the considered sky areas.
We present here preliminary results of the first APS local 
analysis on the whole sky (\cite{Laportaetal2006b,Laportaphd}), 
previously anticipated 
in \cite{Buriganaetal2005}. 
By using {\tt HEALPix} tessellation, 
we divide the sky in patches of a size of $\simeq 14.7^\circ$ which 
allows to accurately compute the APSs~\footnote{We derived 
the $T$ and $PI$ APSs both
passing through the computation of the angular correlation function
and directly with {\tt anafast} to check the reliability of
the found results.}
at $\ell \gsim 60$ (while
angular resolution limits the analysis to $\ell \lsim 300$).
We then fit the APSs
with the same parametrization of Sect.~4.1,
by previously removing bright discrete sources in total intensity
as in Sect.~4.1. The result
is shown in Fig.~\ref{analysis_allsky}.
Clearly, both the APS amplitude (reported here at a reference
multipole $\ell = 100$) and the spectral index exibit
very different patterns in total intensity and polarization,
much more related to the Galactic latitude in the former case.
In total intensity we find a certain APS steepening for increasing power
while this kind of correlation does not appear in polarization, or,
in other words, the region at relatively higher power
and flatter APS in the $C_{\ell=100}$-$\alpha$ plane
is more populated in polarization than in total intensity.

 \begin{figure}
\hskip -0.5cm
   \includegraphics[width=8.cm,height=8cm]{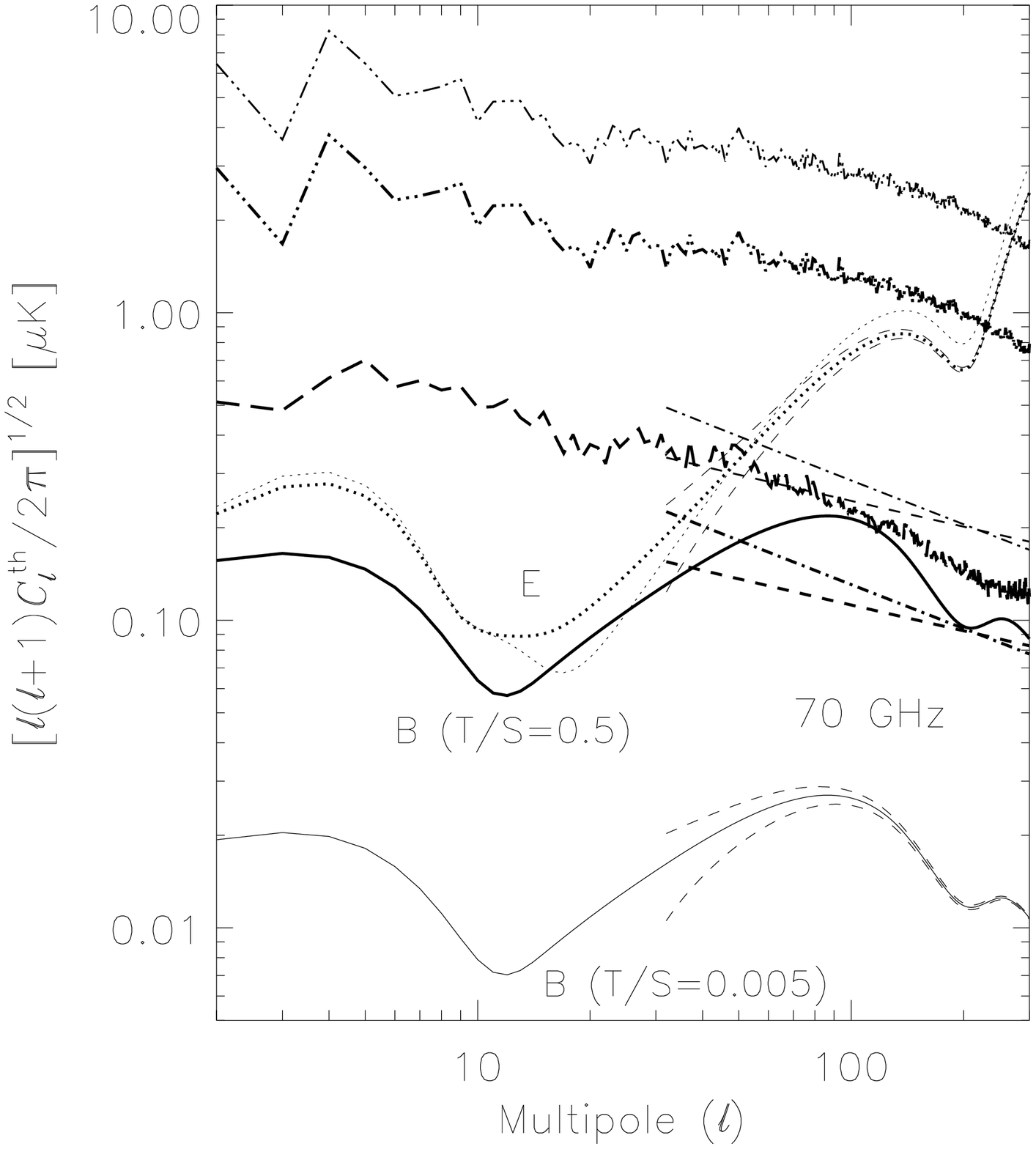}
   \includegraphics[width=8.cm,height=8cm]{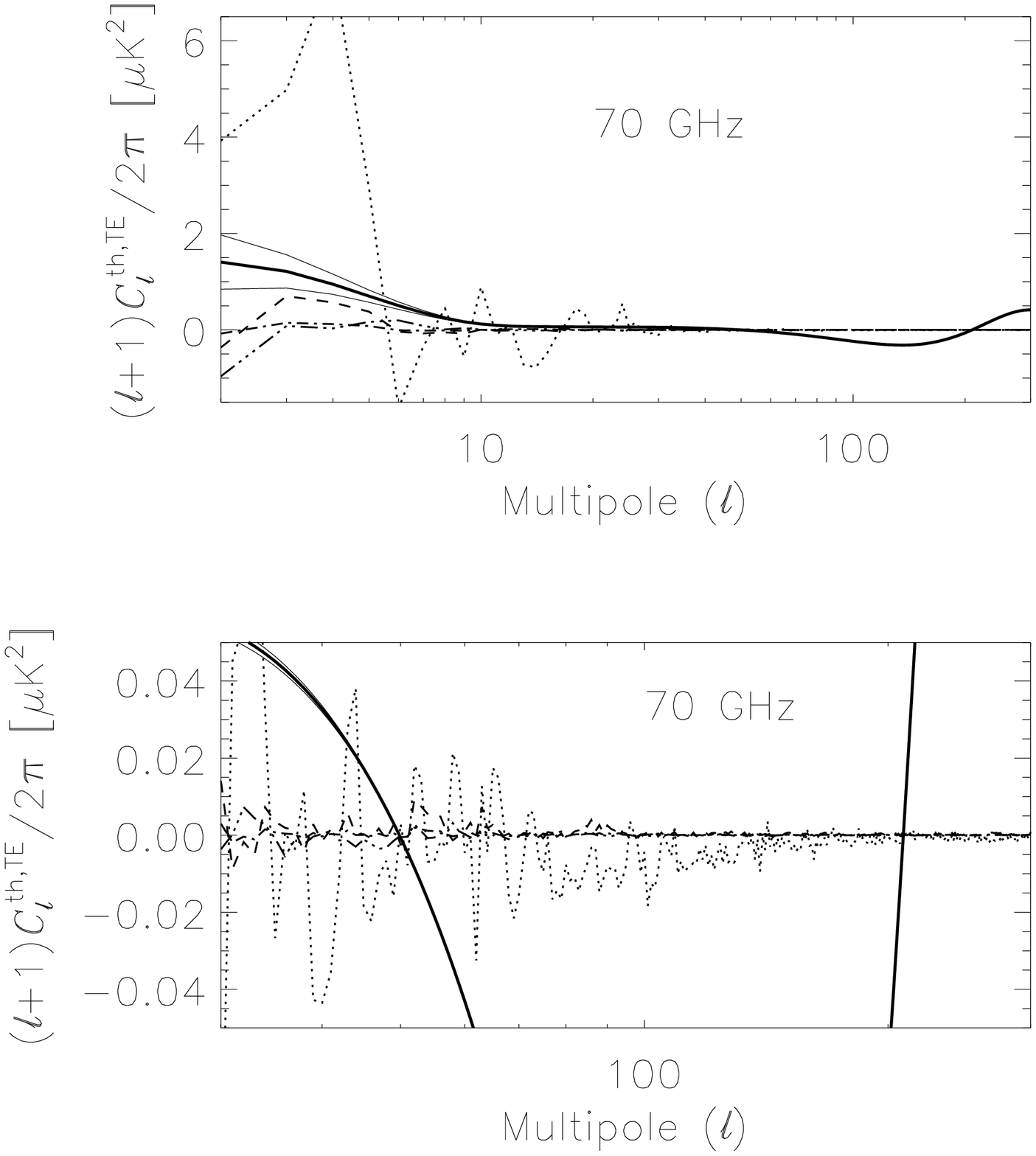}
   \caption{Comparison between the $E$ and $B$ modes (right panel) and the $TE$ mode (left panels) 
of the CMB anisotropy
at 70 GHz and of the Galactic polarized synchrotron foreground.
    Dotted and solid lines: 
    CMB $E$ and $B$ mode for two different tensor to scalar ratios 
(thin line $T/S=0.005$, thick line $T/S=0.5$). 
    Dashed lines: 
    uncertainty associated to the cosmic and sampling variance
    for a sky coverage of $\sim 3\%$
    and a binning of 10\% in $\ell$. 
    Three dots-dashes: 
    average of the $E$ and $B$ modes extrapolated 
    from the whole DRAO survey for 
   the spectral indices $\beta = -2.8$ and $-3$, respectively. 
    The {\tt anafast} results have been divided by the window 
    function 
    to correct for beam smoothing.
    Thick long dashed line: the same as above but with a 
    mask at $|b| \le 20$ for $\beta=-3$.
    Dashed (dot-dashed) power law: best fit result 
    corresponding to Patch~B (resp.~C), rescaled in frequency as above.
The CMB $TE$ mode (thick solid line) 
is for $T/S=0.005$; the corresponding 
cosmic variance (without binning in $\ell$)
identified by the thin solid lines
assumes all-sky coverage. 
 Dots:  extrapolated DRAO $TE$ mode 
for a spectral index $\beta=-3$.
Dashes (dot-dashes): 
as above but masking the region 
at $|b| \le 5^\circ$ and adopting
 $\beta = -2.8$ ($\beta =-3$).
Three dots-dashes: as above but excluding the region 
at $|b| \le 20^\circ$  ($\beta = -2.8$).
Top and bottom panels are identical but with a different choice
of the multipole and power range, for a better view of the results. 
The results are in terms of thermodynamic temperature,
as usual in CMB studies. Adapted from \cite{laportaetal06}.
}
  \label{fig:eb}
 \end{figure}

\section{Comparison with CMB polarization anisotropy}

The accurate measure of the $E$ mode is of particular relevance 
for breaking the existing degeneracy in cosmological parameter estimation 
when only temperature anisotropy data are available (e.g. 
\cite{bond95,efstat99}). 
The detection of the primordial $B$ mode is of fundamental importance
for testing the existence of a stochastic cosmological 
background of gravitational waves
(e.g. \cite{knox94}).
The frequency range $60-80$~GHz seems
the less contaminated by Galactic synchrotron and dust foregrounds in both 
temperature (\cite{bennett03}) and polarization (\cite{page06})
at angular scales $\gsim 1^\circ$.

We have extrapolated the 
APS for the $E$ and $B$ modes 
derived from the DRAO survey analysis
to 70~GHz and compared it with the APS of the CMB polarization 
anisotropy
for a $\Lambda$CDM model including scalar and tensor 
perturbations compatible with the recent WMAP 3~years 
results\footnote{http://lambda.gsfc.nasa.gov/product/map/current/
} 
(\cite{spergel06}). 
The results of this comparison (\cite{laportaetal06}) are displayed in Fig.~\ref{fig:eb}
for two choices of the temperature spectral index 
($\beta = -2.8, -3$). 
The APS extrapolated from the entire DRAO survey is also shown. It
exceeds the CMB $E$ mode even at the
peak at $\ell \sim 100$ for $\beta = -2.8$. For 
$\beta = -3$ the two APSs are almost comparable.
Fig.~\ref{fig:eb} shows also the APS extrapolated  
from the DRAO survey excluding the region $|b| \le 20^{\circ}$ 
and adopting $\beta = -3$.
Such a sky mask 
reduces the Galactic APS below the CMB $E$ mode for 
$\ell \gsim 50$.
In this case the CMB $B$ mode
peak at $\ell \sim 100$ is comparable to (or exceeds) the synchrotron signal 
for tensor to scalar ratios $T/S \gsim 0.5$.
For the three patches
and $T/S \gsim 0.5$ 
the cosmological $B$ mode peak at $\ell \sim 100$ is comparable to (or exceeds) the Galactic 
synchrotron signal extrapolated with $\beta \simeq -2.8$, while 
it is larger by a factor $\gsim 2$ 
(in terms of $\sqrt{C_\ell}$)
for $\beta \simeq -3$. Furthermore,
a separation of the Galactic synchrotron polarized signal from the CMB signal
with an accuracy of $\sim 5-10$\% (in terms of $\sqrt{C_\ell}$) would allow 
to detect the CMB $B$ mode peak at $\ell \sim 100$ even for 
$T/S \sim 0.005$ if $\beta \simeq -3$. 
 Similar results for the 
 detection of the B mode peak at $\ell \sim 100$ have been obtained 
  by \cite{carretti06} based on 1.4~GHz Effelsberg polarization 
 observations of a small region with
 exceptionally low Galactic foreground contamination, though
 at $\ell \sim {\rm few} \times 100$ the CMB $B$ mode is expected to be
 dominated by the $B$ mode generated by lensing 
 (\cite{zaldarriaga98}).
We note also that for a sky coverage comparable with those 
of the considered patches
($\sim 3$\%) the cosmic and sampling variance does not significantly
limit the accuracy of the CMB mode recovery 
at $\ell \gsim {\rm some} \times 10$.\\

The CMB $TE$ correlation sets constrains
on the reionization history from the power at low multipoles 
and on the nature of primordial fluctuations 
from the series of peaks and antipeaks. 
In Fig.~\ref{fig:eb}
we compare the $TE$ mode APS of the Galactic 
 emission extrapolated at 70~GHz with the CMB one.
The APS of the whole DRAO survey indicates 
a significant Galactic contamination at $\ell \lsim {\rm few} \times 10$ 
 even for $\beta \simeq  -3$. 
The use of a proper Galactic mask (e.g. excluding
regions with $|b| \le 5^\circ$ for $\beta\sim-3$
or $|b| \le 20^\circ$ for $\beta\sim-2.8$) 
largely reduces the Galactic foreground
contamination even at low multipoles.
For the three cold patches the Galactic $TE$ mode 
is fully negligible compared to the CMB one independently 
of the adopted $\beta$. 
The $TE$ mode antipeak at $\ell \sim 150$ turns out to be 
very weakly affected by Galactic synchrotron contamination in all
cases. These considerations (\cite{laportaetal06}) further support the reliability of 
the $TE$ mode measure performed by WMAP and of the derived conclusions
on the epoch of reionization and  
of a dominance of adiabatic perturbations and existence
of superhorizon temperature fluctuations at decoupling
(\cite{kogut03,page06})  
confirming the inflationary paradigm (\cite{peiris03}).
 
\begin{figure}
   \centering
\includegraphics[width=14.cm,height=8.cm]{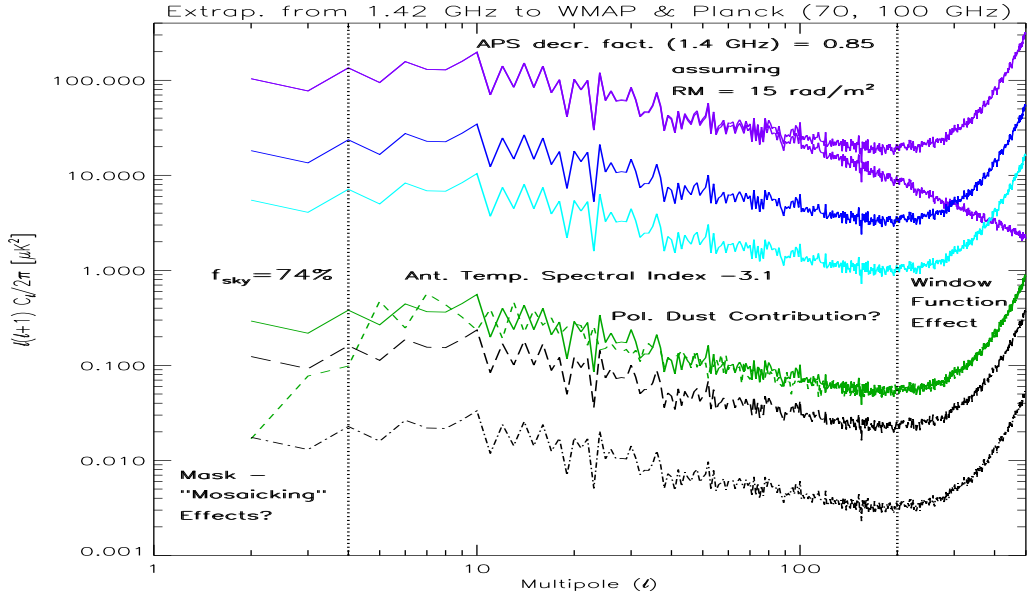}
   \caption{Galactic synchrotron APSs 
($E$ mode, except for (green) short dashes that refer to $B$ mode at 60.3~GHz, 
for comparison)
of the Galactic synchrotron emission extrapolated 
to $\nu/$~GHz~=~22.8, 30.3, 36.9, 60.3, 70, 100 
(from top to bottom) from the 1.4~GHz all-sky survey 
applying the WMAP polarization mask. For a better comparison with 
Fig.~17 of \cite{page06},
the APSs have been divided by the window function to account for beam smoothing and
partially correct the power at $\ell \sim 100-200$,
but the power increasing at $\ell \gsim 200$ is clearly non-physical
(at 22.8~GHz the smoothed APS is also shown). }
  \label{fig:eb_allsky}
 \end{figure}

The comparison of WMAP 3yr results on polarized foregrounds
with those derived from radio surveys is of particular 
interest in the case of all-sky coverage.
For a better comparison, we applied to the all-sky 1.4~GHz 
polarization map the same 
polarization mask considered by the WMAP team, which excludes
$\simeq 24$\% of sky (mainly the Galactic plane and the NPS). 
In this comparison, we exploit a first order correction for
Faraday depolarization by increasing the 1.4~GHz polarized APS 
of a factor $1/0.85$ corresponding to a typical value of 
$RM \sim 15$~rad/m$^2$, as inferred from Sect.~3, before applying
a simple power law frequency extrapolation with constant $\beta$
to WMAP and {\sc Planck}~\footnote{http://www.rssd.esa.int/planck} frequencies.
The result is shown in Fig.~\ref{fig:eb_allsky}. 
It can be easily compared with 
Fig.~17 of \cite{page06}. Clearly, radio data are more sensitive
to synchrotron emission. This opportunity reflects into the more accurate and regular
APS shape, particularly at $\ell \gsim 10$ where the intercalibration 
of the various sky areas is less critical.
At $\nu < 40$~GHz there is a good agreement
between the average APSs extrapolated from radio surveys and those derived
with WMAP, which includes the overall contribution of polarized foregrounds.
In spite of the difficulty to fully
understand the complexity of ISM structure and depolarization effects 
because of the role of various physical processes
(turbulences in both the Galactic magnetic field and the 
ISM electron density depolarizing the diffuse synchrotron
emission (see \cite{hav04}), 
foreground magneto-ionic structures (Faraday screens) 
modulating the background synchrotron diffuse emission 
(see \cite{Wolleben&Reich04,Uyaniker03}), probably more relevant 
at smaller scales),
this agreement supports the idea that the bulk of 
the correlation properties of the
microwave synchrotron polarized emission at 
$\ell \lsim 300$ can be 
mapped on the basis of radio data resorting to 
relatively simple Faraday depolarization arguments 
and frequency rescaling. 
On the contrary, in the V band the WMAP foreground APS 
is about a factor
of two above that extrapolated from radio surveys. 
If not explained by an unlikely significant frequency flattening of the 
synchrotron polarized emission at $\nu \gsim 40$~GHz
with respect to the behaviour at $\nu \lsim 40$~GHz,
this discrepancy can be only explained in terms of 
additional polarized contributions
in the V band (with a global power similar to that of the diffuse 
synchrotron emission), to be probably ascribed to polarized 
dust emission.

\vskip 0.5cm
\noindent
{\bf Acknowledgements --}
We warmly thank R.~Beck, P.~Leahy, T.A.T. Spoelstra, L.~Toffolatti, M.~Tucci, 
and R.~Wielebinski for useful discussions.
Some of the results in this paper have been derived using {\tt HEALPix}
(\cite{gorski05}). The use of the {\tt CMBFAST} code
(version 4.5.1) is acknowledged. L.L.P. was supported for this research
through a stipend from
the International Max Planck Research School (IMPRS)
for Radio and
Infrared Astronomy at the universities of Bonn and
Cologne.


\begin{thebibliography}{99}

  \bibitem{bacci01}
   Baccigalupi, C., Burigana, C., Perrotta, F., et al., 2001, 
   A\&A, 372, 8

    \bibitem{bennett03}
     Bennett, C.~L., Hill, R.~S., Hinshaw, G., et al., 2003,
     ApJS, 148, 97	

  \bibitem{bingham67} 
   Bingham, R.~G., 1967, 
   MNRAS, 137, 157
   
  \bibitem{bond95}
  Bond, J.~R., Davis, R.~L., Steinhardt, P.~J., 1995
  ApL\&C, 32, 53

  \bibitem{spo76}
   Brouw, W.~N., Spoelstra, T.~A.~T., 1976, 
   A\&AS, 26, 129

  \bibitem{buriganalaporta02}
   Burigana, C., La~Porta, L., 2002, 
   in Astrophysical Polarized Background,
   AIP Conf. Proc. 609, 54, 
   astro-ph/0202439

  \bibitem{Buriganaetal2005}
   Burigana, C., La Porta, L., Reich, P., Reich, W., 2005, 
talk at the Workshop ``Polarisation 2005 - Sky polarisation at
far-infrared to radio wavelengths:
The Galactic Screen before the Cosmic Microwave Background'',
University Paris-XI campus, September 12-15, 2005

  \bibitem{Buriganaetal2006}
   Burigana, C., La Porta, L., Reich, P., Reich, W., 2006, AN, 327, No. 5/6, 491


  \bibitem{burn66}
   Burn, B.~J.~1966, MNRAS, 133, 67

   \bibitem{carretti06}    
    Carretti, E., Poppi, S., Reich, W., et al., 2006,
    MNRAS, 367, 132

   \bibitem{chepurnov}
   Chepurnov, A.~V., 1999, Astron. Astrophys. Trans., 17, 281

   \bibitem{cho_lazarian}
    Cho, J., Lazarian, A., 2002, ApJ, 575, L63

   \bibitem{dineen}
   Dineen, P., Coles, P., 2005,
   MNRAS, 362, 403

  \bibitem{efstat99}  
  Efstathiou, G., Bond, J.~R., 1999,
  MNRAS, 304, 75	

  \bibitem{egger95}
   Egger, R.~J., Aschenbach, B., 1995, 
   A\&A, 294, L25 

  \bibitem{GardnerWhiteoak66}
   Gardner, F.~F., Whiteoak, J.~B., 1966, 
   ARA\&A, 4, 245

  \bibitem{ginzburg65}
   Ginzburg, V.~L., Syrovatski, S.~I., 1965,
   ARA\&A, 3, 297 

  \bibitem{gorski05} 
   G\'orski, K.~M., Hivon, E., Banday, A.~J., et al., 2005,
   ApJ, 622, 759 

  \bibitem{haslam74} Haslam, C.~G.~T., 1974,
           A\&AS,15,333

  \bibitem{hav04} 
   Haverkorn, M., Katgert, P., de Bruyn, A.~G., 2004,
   A\&A, 427, 549

   \bibitem{johnston}
    Johnston-Hollitt, M., Hollitt, C.~P., Ekers, R.~D., 2004,
    in The Magnetized Interstellar Medium,
   ed. B.~Uyaniker, W.~Reich, R.~Wielebinski
    (Copernicus GmbH), 13,
    http://www.mpifr-bonn.mpg.de/div/konti/antalya/contrib.html    
   
  \bibitem{kamion}
   Kamionkowski, M., Kosowsky, A., Stebbins, A., 1997, 
   Phys.Rev. D, 55, 7368

  \bibitem{knox94}   
   Knox, L., Turner, M.~S., 1994,
   PhRvL, 73, 3347

  \bibitem{kogut03}
   Kogut, A.,~2003,
   NewAR, 47, 977	

  \bibitem{Kosowsky99}
   Kosowsky, A., 1999, 
   NewAR, 43, 157 

  \bibitem{Laportaphd}
  La Porta, L., 2006, {\it PhD Thesis}, Bonn University, Germany, in prep.

  \bibitem{laportaburigana06}
   La Porta, L., Burigana, C., 2006, A\&A, in press, astro-ph/0601371

  \bibitem{laportaetal06}
   La Porta, L., Burigana, C., Reich, W., Reich, P., 2006, A\&AL, in press,
  astro-ph/0606xxx

  \bibitem{Laportaetal2006b}
  La Porta, L., Burigana, C., Reich, W., Reich, P., 2006, in prep.

   \bibitem{manchester98}
   Manchester, R.~N., Han, J.~L., Qiao, G.~J., 1998,
   MNRAS, 295, 280

  \bibitem{Massardietal2006}
   Massardi, M., Gonzalez-Nuevo, J., De Zotti, G., 2006, this conference

   \bibitem{mesa02}
   Mesa, D., Baccigalupi, C., De Zotti, G., et al., 2002, 
   A\&A, 396, 463

  \bibitem{page06}
   Page, L., Hinshaw, G., Komatsu, E., et al., 2006, ApJ, submitted,
   astro-ph/0603450


  \bibitem{peebles} 
   Peebles, P.~J.~E., 1993, 
   {\it Principles of Physical Cosmology},
   Princeton University Press

  \bibitem{peiris03}
    Peiris, H.~V., Komatsu, E., Verde, L., et al., 2003,
   ApJS, 148, 213


  \bibitem{platania98}
   Platania, P., Bensadoun, M., Bersanelli, M., et al., 1998,
   ApJ, 505, 2, 473


  \bibitem{prandoni01}
   Prandoni, I., Gregorini, L., Parma, P., et al., 2001, A\&A, 365, 392

  \bibitem{reich86}
   Reich, P., Reich, W., 1986,
   A\&AS, 63, 205 

  \bibitem{reich88}
   Reich, P., Reich, W., 1988,
   A\&AS, 74, 7 

  \bibitem{testori01} 
   Reich, P., Testori, J.~C., Reich, W., 2001,  
   A\&A, 376, 861

  \bibitem{reich82}
   Reich, W., 1982,
   A\&AS, 48, 219 

  \bibitem{reich06}
   Reich, W., 2006,
    in Review Book ``Cosmic Polarization'', ed. R.~Fabbri, Publisher: Research Signpost,
    astro-ph/0603465

  \bibitem{salter83} 
   Salter, C.~J., 1983, 
   BASI, 11, 1

  \bibitem{seljak97}
  Seljak, U., Zaldarriaga, M., 1997, 
Phys. Rev. Lett., 
78, 2054

  \bibitem{sokoloff}
   Sokoloff, D.~D., Bykov, A.~A., Shukurov, A., et al., 1998,
   MNRAS, 299, 1985
  

  \bibitem{spergel06}
  Spergel, D.~N., Bean, R., Dore, O., et al., 2006, ApJ, submitted, astro-ph/0603449

  \bibitem{spoelstra84}
   Spoelstra, T.~A.~T., 1984, 
   A\&AS, 135, 238 

  \bibitem{testori03} 
   Testori, J.~C., Reich, P., \& Reich, W. 2003,
   in The Magnetized Interstellar Medium,
   ed. B.~Uyaniker, W.~Reich, and R.~Wielebinski
   (Katlenburg-Lindau: Copernicus GmbH), 57,
   http://www.mpifr-bonn.mpg.de/div/konti/antalya/contrib.html
   
 \bibitem{toffolatti98}   
  Toffolatti, L., Arg\"ueso~G\'omez, F., De Zotti, G., et~al., 1998,
  MNRAS, 297, 117

 \bibitem{tucci04}   
  Tucci, M., Mart\'inez-Gonz\'alez, E., Toffolatti, L., et al.~2004,
  MNRAS, 349, 1267


  \bibitem{Uyaniker03} 
   Uyaniker, B., 2003, 
   in The Magnetized Interstellar Medium,
   ed. B.~Uyaniker, W.~Reich, R.~Wielebinski
   (Katlenburg-Lindau: Copernicus GmbH), 71, 
   http://www.mpifr-bonn.mpg.de/div/konti/antalya/contrib.html

  \bibitem{VinyajkinRazin2002} 
   Vinyajkin, E.~N., Razin, V.~A., 2002,
   in Astrophysical Polarized Background,
   AIP Conf. Proc. 609, 26 

  \bibitem{wollebenPhD} 
   Wolleben, M., 2005, {\it PhD Thesis}, Bonn University, Germany 

  \bibitem{wolleben06}
   Wolleben, M., Landecker, T.~L., Reich, W., Wielebinski, R., 2006,
   A\&A, 448, 411  

  \bibitem{Wolleben&Reich04}
   Wolleben, M., Reich, W., 2004, 
   A\&A, 427, 537

  \bibitem{zald}
   Zaldarriaga, M., 2001, Phys. Rev. D, 64, 15 
 
  \bibitem{zaldarriaga98}           
   Zaldarriaga, M., Seljak, U., 1998,   
   Phys. Rev. D, 58, 3003
   
\end{thebibliography}
\end{document}